\documentclass[galaxies,article,accept,pdftex,moreauthors]{Definitions/mdpi}
\newcommand{\bhspin}{a_*}

\firstpage{1}
\pubvolume{1}
\issuenum{1}
\articlenumber{0}
\pubyear{2026}
\copyrightyear{2026}
\datereceived{ }
\daterevised{ } % Comment out if no revised date
\dateaccepted{ }
\datepublished{ }
\Title{Signatures of Black Hole Spin in Horizon-Scale Polarimetry}

\Author{George N.~Wong $^{1,2}$\orcidA{}*, Daniel C.~M.~Palumbo $^{3,4}$\orcidB{}, Zachary Gelles $^{2,5}$\orcidC{}, and Andrew Chael $^{6}$\orcidD{}}

\AuthorNames{George N. Wong, Daniel C. M. Palumbo, Zachary Gelles, and Andrew Chael}

\address{%
$^{1}$ \quad School of Natural Sciences, Institute for Advanced Study, 1 Einstein Drive, Princeton, NJ 08540, USA\\
$^{2}$ \quad Princeton Gravity Initiative, Princeton University, Princeton, NJ 08544, USA\\
$^{3}$ \quad Black Hole Initiative at Harvard University, 20 Garden Street, Cambridge, MA 02138, USA\\
$^{4}$ \quad Center for Astrophysics | Harvard \& Smithsonian, 60 Garden Street, Cambridge, MA 02138, USA\\
$^{5}$ \quad Department of Physics, Princeton University, Princeton, NJ 08540, USA\\
$^{6}$ \quad Niels Bohr International Academy, Niels Bohr Institute, Blegdamsvej 17, DK-2100 Copenhagen \O, Denmark
}

\corres{Correspondence: gnwong@ias.edu}

\abstract{The angular momentum of a black hole, usually expressed in terms of a dimensionless ``spin,'' both shapes the strong-field spacetime and provides a reservoir of rotational energy that can be exchanged with surrounding plasma. Very long baseline interferometry (VLBI) has now begun to resolve polarized emission on event-horizon scales. We distinguish polarimetric signatures of spin arising primarily from photon propagation in the Kerr spacetime from those mediated by horizon-threading electromagnetic fields and magnetized plasma dynamics. We trace the inference from VLBI correlations through Stokes images and compact summary statistics to constraints on the source and, ultimately, on spin. Within this framework, we review diagnostics linked to horizon regularity and magnetic-field winding, magnetically arrested accretion, electromagnetic energy extraction, jet-base and light-cylinder structure, and horizon and photon-ring polarization. Current Event Horizon Telescope observations constrain magnetic-field geometry, variability, source orientation, magnetic flux state, and aspects of the disk--jet connection more robustly than they constrain spin magnitude or sense. Future observing and modeling programs should prioritize combinations of polarimetric diagnostics with complementary systematics and test whether a common spin-dependent interpretation is supported across independent data products and plausible source models.}

\keyword{black holes (162), Kerr black holes (886), polarimetry (1278), accretion (14), magnetic fields (994), relativistic jets (1390), event horizons (479), radiative transfer (1335), very long baseline interferometry (1769)}

\begin{document}

\section{Introduction}

In general relativity, an isolated stationary black hole is completely described by its mass, electric charge, and angular momentum \citep{kerr_1963_gravfield,johannsen_2010_nohairimages,chrusciel_2012_uniqueness}. Mass sets the length and time scales of the near-horizon problem, and for the best horizon-scale targets, it can be constrained independently of the radio polarimetric image. In realistic astrophysical environments, ambient plasma rapidly neutralizes any net charge \citep{eardley_1975_astroproc}. This article focuses on the final parameter, the black hole angular momentum, which is usually expressed in terms of a dimensionless spin parameter $\bhspin \equiv Jc/GM^2$. Unless otherwise specified, we use the signed value of $\bhspin$ to denote prograde (retrograde) alignment with the accretion flow as positive (negative) spin and treat the spin-axis orientation on the sky as a separate quantity. Spin affects how matter moves near the event horizon, how efficiently accretion can release energy, and whether large-scale magnetic fields can extract rotational energy \citep{bardeen_1972_kerr,blandford_1977_bz,znajek_1977_condition,mckinney_2004_kerrjets,mckinney_2006_jets,tchekhovskoy_2011_mad}. Because angular momentum is accumulated and redistributed during accretion and mergers, the magnitudes and orientations of black hole spins also carry information about black hole assembly histories and demographics \citep{king_2006_chaoticaccretion,volonteri_2010_bhgrowth,barausse_2012_bhspins,sesana_2013_insightsbh,reynolds_2021_spinreview}. Spin is thus astrophysically important because it acts as a spacetime parameter, as a reservoir of rotational energy that can shape nearby plasma, magnetic fields, and outflows, and as a tracer of the evolution of the universe.

Progress over the past decade has brought black hole spin within observational reach across a wide range of masses and environments. The Event Horizon Telescope (EHT) has resolved electromagnetic emission on event horizon scales in M87* and Sgr A*, producing images of the plasma immediately surrounding the two supermassive black holes with the largest apparent angular sizes on the sky \citep{eht_m87_1,eht_m87_5,eht_m87_7,eht_m87_8,eht_m87_2018_1,eht_m87_20172021,eht_sgra_1,eht_sgra_5,eht_sgra_7,eht_sgra_8}. Ground-based gravitational wave detectors have constrained the masses and spins of stellar-mass black holes during dynamical mergers, and pulsar timing arrays have begun to probe the gravitational wave background likely produced by massive black hole binaries \citep{abbott_2023_gwtc3,lvk_2023_population,biscoveanu_2026_gwspins,nanograv_2023_15yr,nanograv_2023_gwbackground,epta_inpta_2023_gwb,epta_inpta_2023_implications,ppta_2023_gwb,ipta_2024_comparing}. X-ray reflection spectroscopy, continuum fitting, timing and reverberation analyses, and X-ray polarimetry can provide information about spin through unresolved radiation from more radiatively efficient accretion flows \citep{brenneman_2013_spinreview,mcclintock_2014_continuum,miller_2009_relativisticxray,bambi_2021_xrayspin,siskreynes_2026_xrayspin,reynolds_2021_spinreview,weisskopf_2022_ixpe}. These measurements connect spin across the black hole mass spectrum and across regimes ranging from dynamical coalescence to approximately stationary accretion. Horizon-scale polarimetry enters this broader program by resolving the strong-field regions around nearby isolated supermassive black holes.

This review is organized around a physical question: how is black hole spin encoded in horizon-scale radio polarimetric images of supermassive black holes? Unfortunately, spin cannot be measured directly given a polarimetric image. Rather, as a parameter of the Kerr spacetime and a reservoir of rotational energy, it shapes photon propagation, horizon boundary conditions, and the magnetized plasma dynamics that produce the polarized emission. Spin constraints therefore depend on models that connect polarization observables to the Kerr geometry and to the accretion flow. To relate polarimetric data to spin, we draw on work using polarization to probe black-hole accretion and outflow physics, from spatially resolved black-hole polarimetry and general relativistic magnetohydrodynamics (GRMHD) predictions for polarization signatures to emission models for disks, coronae, and jets. Several recent reviews have discussed complementary aspects of this modeling and inference problem, including how resolved polarimetric observables constrain accretion flow models and how polarization signatures are extracted from numerical simulations \citep{ricarte_2023_resolvedpol,fragile_2026_grmhdpol}. Here we review the physical mechanisms connecting spin to horizon-scale and long-baseline radio polarimetric observables, with an emphasis on resolved synchrotron polarization, compact image descriptors, jet-base polarimetry, and horizon, photon-ring, and subring signatures.

The EHT provides a contemporary observational context for this question. In total intensity, M87* and Sgr A* show compact ring-like emission on scales comparable to the gravitational radius, and the measured ring diameters are consistent with expectations for gravitationally lensed emission around black holes of the independently measured masses \citep{eht_m87_1,eht_sgra_1,eht_sgra_6}. Features such as ring diameter, width, and asymmetry leave many spin-related source properties underdetermined, however, because similar images can be produced by different emission geometries, plasma thermodynamics, magnetic flux states, viewing inclinations, and spin values once sparse interferometric sampling, finite angular resolution, and source variability are taken into account \citep{eht_m87_5,eht_m87_2018_2,eht_sgra_5,wong_2022_patoka,chael_2025_radsurvey,dhruv_2025_surveyv3}.

\begin{figure}[H]
\begin{adjustwidth}{-\extralength}{0cm}
\centering
\includegraphics[width=1.3\textwidth]{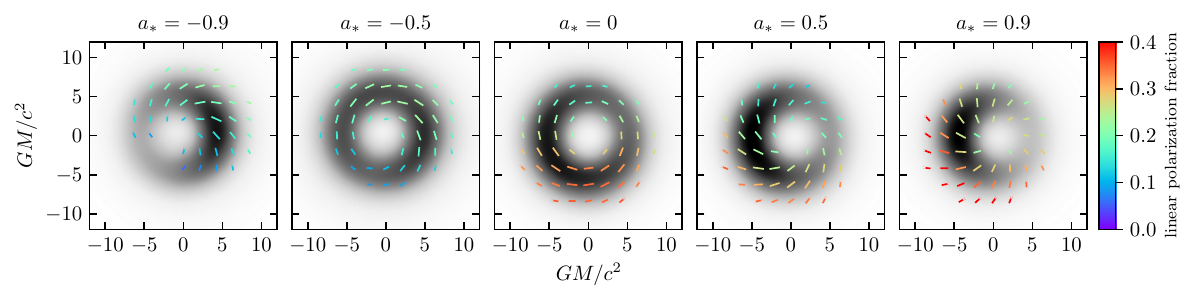}
\end{adjustwidth}
\caption{Representative ray-traced GRMHD images of highly magnetized accretion flows for five different spins, blurred with a $15\,\mu{\rm as}$ Gaussian to simulate EHT resolution. The grayscale background encodes total intensity, tick directions denote the EVPA, and tick color gives the local linear-polarization fraction. The five models hint at a relationship between the polarimetric observables and the black hole spin and illustrate the image-plane structure that motivates compact polarization descriptors. The simulations were produced using the {\texttt{AthenaK} code \citep{stone_2026_athenak}, first described in \citet{wong_2025_mixing}, and imaged for M87*-like parameters with $r_{\rm low} = 1$ and $r_{\rm high} = 40$. See \citet{wong_2022_patoka} for more detail.}}
\label{fig:survey_spin_gallery}
\end{figure}

But polarimetry adds information that is not present in total intensity. The EHT polarized images of M87* and Sgr A* show organized structure near the event horizon, and current interpretations use these data to constrain magnetic-field geometry, Faraday depth, and plasma conditions near the black hole \citep{eht_m87_7,eht_m87_8,eht_m87_9,eht_sgra_7,eht_sgra_8}. For synchrotron emission, linear polarization depends on the projected magnetic-field structure, optical depth, and relativistic transport of the polarization basis between the source and the observer. Faraday rotation changes the observed electric vector position angle (EVPA), while Faraday conversion transfers power between the linear and circular polarization modes. Although these effects are harder to measure and interpret than total intensity, they are sensitive to magnetic polarity, electron temperature, composition, and the distribution of plasma along the line of sight. In simulations, the aggregate of this information has been demonstrated to imprint clear geometric signatures onto images. Figure~\ref{fig:survey_spin_gallery} shows a representative set of ray-traced GRMHD images across a range of black hole spins, illustrating possible spin-sensitive differences in the structure and magnitude of the linear polarization.

Horizon-scale polarimetric data therefore constrain the magnetized source more directly than the Kerr spin. Spin information enters the observed polarization through photon propagation in the Kerr spacetime, electromagnetic boundary conditions at the horizon, and the influence of spin on the magnetized plasma dynamics. These routes differ in observational accessibility and in the extent to which their spin dependence can be reproduced by changes in the plasma, emission geometry, or polarized transfer. Interpreting the spin-sensitive features in a polarimetric image requires separating the source physics from the relativistic mapping of that source to the observer. On the source side, orbital shear, frame dragging, inflow and outflow structure, and the boundary conditions supplied by the disk and horizon help set the geometry and winding of the magnetic field in the emitting region \citep{narayan_2003_mad,tchekhovskoy_2011_mad,mckinney_2012_mad,begelman_2022_whatmad}. Large-scale magnetic fields may also connect the compact emission region to the jet and, when they thread the horizon, mediate electromagnetic energy extraction through the Blandford--Znajek (BZ) mechanism \citep{blandford_1977_bz,znajek_1977_condition,eht_m87_5,chael_2023_bhp1}.

On the mapping side, lensing and parallel transport organize the direct image of the emitting plasma, near-horizon structure, and higher-order lensed subimages. The direct image is produced by photons that reach the observer without passing around the far side of the black hole, while higher-order subimages are produced by photons that pass close to unstable photon trajectories before escaping. These higher-order components form the photon-ring structure. Individual components corresponding to a particular integer number of half-orbits are often called subrings, and they accumulate near the critical curve, the limiting image-plane curve approached by photons that remain near unstable bound trajectories for many orbits. The locations and polarization patterns in the subrings can be tied directly to lensing and parallel transport in the Kerr spacetime. Together, these source and mapping effects can shape observable signals such as the handedness and coherence of the polarization pattern, the orientation of the compact polarized emission relative to the larger-scale jet, the locations and polarization patterns of higher-order lensed subrings, and the time dependence of the emitting plasma \citep{palumbo_2020_beta2,wong_2026_bhp2,gelles_2025_spinjetpol,gelles_2026_offaxisjetpolspin,himwich_2020_universalpol,palumbo_2022_photonringbeta2}. Although none of these mechanisms produces a one-to-one map from a polarized image to $\bhspin$, together they make polarimetry a natural place to look for spin-sensitive information.

The remainder of this article follows a progression from measurement to interpretation. In Section~\ref{sec:observables}, we describe what horizon-scale polarimetry measures and how polarized images and interferometric observables are related to the emitting plasma. Then, in Section~\ref{sec:connection}, we discuss the main mechanisms by which spin-dependent information can be encoded in polarimetric observables, beginning with compact EHT emission-region observables, then jet-base polarimetry and the disk--jet connection, and finally near-horizon, photon-ring, and long-baseline signatures. We summarize the current status of spin inference from horizon-scale and related polarimetry in Section~\ref{sec:current}, separating quantities that polarimetry can measure directly from quantities one may wish to infer, including spin magnitude, spin sign, spin-axis orientation, electromagnetic energy extraction, magnetic flux state, and plasma conditions. Section~\ref{sec:outlook} closes by identifying the main limitations and the observations and models most likely to make future spin claims more robust.

\section{Polarimetric Observables on Horizon Scales}
\label{sec:observables}

It is useful to separate the problem into three layers: the physical source, the sky brightness distribution (image), and the measured interferometric data. Magnetized plasma near the black hole emits polarized synchrotron radiation, and relativistic radiative transfer maps that emission into a sky brightness distribution in Stokes $I$, $Q$, $U$, and $V$. Very long baseline interferometry (VLBI) then correlates signals recorded by pairs of telescopes, and each telescope pair defines a baseline that samples a Fourier component of the sky brightness distribution. Imaging algorithms, geometric modeling, and compact image summary statistics compress these data into lower-dimensional descriptions. Spin is part of the source model: together with the mass, it specifies the spacetime geometry, while the full set of system parameters determines the plasma dynamics and magnetic field structure whose signatures propagate through the subsequent layers.

\begin{figure}[H]
\begin{adjustwidth}{-\extralength}{0cm}
\centering
\includegraphics[width=1.3\textwidth]{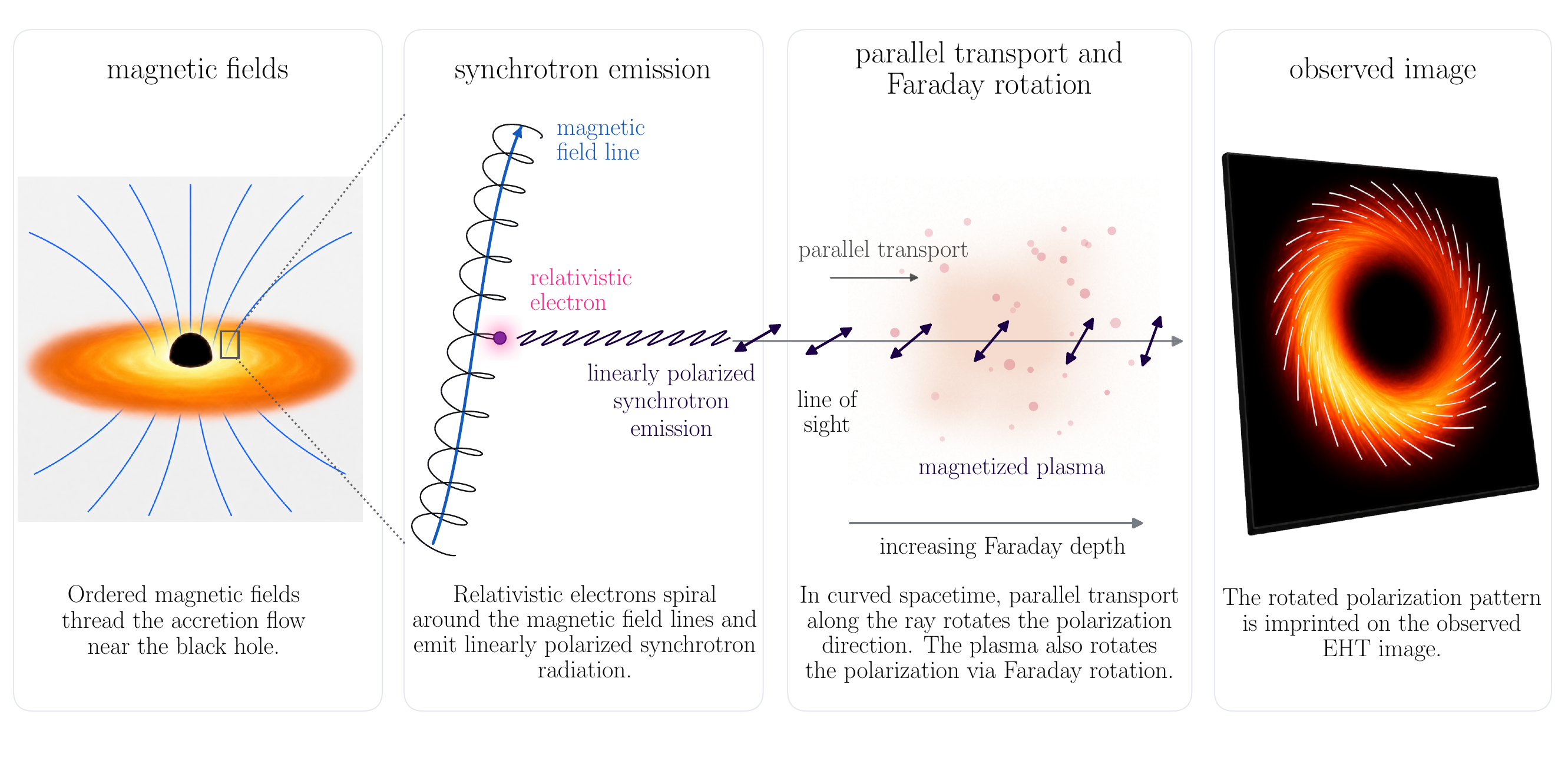}
\end{adjustwidth}
\caption{Schematic path from near-horizon magnetic fields to an observed polarized image. The local synchrotron polarization is determined by the magnetic field geometry in the emitting plasma, but relativistic propagation, parallel transport, and Faraday effects can rotate or depolarize the signal, so the observed EVPA pattern is a processed signature of the source rather than a direct magnetic-field map.
}
\label{fig:sync_cartoon}
\end{figure}

Here we consider millimeter- and submillimeter-wavelength polarimetric observations, which resolve black-hole-scale structure from a few to several hundred gravitational radii, $r_g \equiv GM/c^2$. Although most of the direct image features discussed below are produced within tens of $r_g$ of the black hole, the jet diagnostics can extend farther along the outflow when the measured field geometry remains tied to the central engine of the jet. Millimeter and submillimeter VLBI observations of M87* and Sgr A* probe this angular-resolution regime. At these wavelengths, low-luminosity accretion flows are synchrotron dominated, and synchrotron radiation is naturally polarized \citep{yuan_2014_review,leung_2011_transfercoeffs}. For an ordered, optically thin synchrotron source with an idealized electron distribution, the local emitted linear polarization fraction can approach $\sim70\%$ in the standard power-law case. Lower measured net or resolved image fractions therefore point to depolarization or cancellation from magnetic-field disorder, unresolved EVPA structure, optical depth, Faraday propagation, turbulence, variability, and image averaging. 

In the simplest, optically thin limit, excluding Faraday propagation effects, and assuming a plasma at rest in flat space, the observed EVPA is perpendicular to the projected magnetic field. Horizon-scale images violate all of these simplifications: the emitting plasma moves relativistically, the radiation is lensed and beamed, the polarization basis is parallel transported through curved spacetime, and optical depth and Faraday effects can alter the polarization state before the light reaches the observer. The measured polarization is therefore a joint signature of the emitted radiation, the magnetic field, relativistic propagation, and plasma transfer. Figure \ref{fig:sync_cartoon} provides a visual summary of how the magnetic field geometry, synchrotron emission, and propagation effects combine to produce the observed polarization pattern.

At each position on the image plane, the polarized brightness is described by four Stokes intensities, $\boldsymbol{S}_\nu \equiv (I_\nu,Q_\nu,U_\nu,V_\nu)$. Here $I_\nu$ is the total specific intensity, $Q_\nu$ and $U_\nu$ encode linear polarization relative to an adopted north/east basis on the sky, and $V_\nu$ encodes circular polarization, with $Q_\nu^2+U_\nu^2+V_\nu^2 \le I_\nu^2$. We write the complex linear polarization as $P_\nu\equiv Q_\nu+iU_\nu$, the complex fractional linear polarization as $m\equiv P_\nu/I_\nu$, and the signed fractional circular polarization as $v\equiv V_\nu/I_\nu$. The polarization-fraction magnitudes are therefore
\begin{align}
    |m| &\equiv \frac{\sqrt{Q_\nu^2+U_\nu^2}}{I_\nu}, &
    |v| &\equiv \frac{|V_\nu|}{I_\nu}.
\end{align}

The local quantities $m$, $v$, $|m|$, and $|v|$ are pointwise quantities derived from the Stokes fields, and compressing a resolved image to scalar polarization fractions requires an additional averaging choice. One choice is to first sum the Stokes fluxes over the image and then form a fraction. For positions indexed by $i$, the corresponding image-integrated net fractions are
\begin{align}
    |m|_{\rm net} &\equiv
    \dfrac{\sqrt{\left(\sum\limits_i Q_{\nu,i}\right)^2+\left(\sum\limits_i U_{\nu,i}\right)^2}}{\sum\limits_i I_{\nu,i}}, &
    v_{\rm net} &\equiv \dfrac{\sum\limits_i V_{\nu,i}}{\sum\limits_i I_{\nu,i}}.
\end{align}
The quantity $v_{\rm net}$ is signed, although EHT circular-polarization limits are often quoted as $|v_{\rm net}|$ or, for ALMA measurements that include emission outside the EHT image field, $|v_{\rm int}|$ \citep{eht_m87_7,eht_m87_8,eht_m87_9}. These ``unresolved'' image-integrated quantities are subject to cancellations across the image due to variations in the EVPAs or differences in the sign of Stokes $V$.

Alternatively, one may first evaluate the local polarization magnitudes and then average them over the image. This gives the ``resolved,'' image-averaged polarization-fraction magnitudes
\begin{align}
    \langle |m| \rangle &\equiv
    \dfrac{\sum\limits_i \sqrt{Q_{\nu,i}^2+U_{\nu,i}^2}}{\sum\limits_i I_{\nu,i}}, &
    \langle |v| \rangle &\equiv
    \dfrac{\sum\limits_i |V_{\nu,i}|}{\sum\limits_i I_{\nu,i}}.
\end{align}
EHT analyses commonly report these resolved averages as standard image-level summaries of local polarization because they are useful for comparing reconstructed or smoothed images given a common, reference angular scale. They are not unique or fundamental descriptors of the resolved polarization field, however, since they depend on the image resolution, smoothing, and chosen averaging region.\footnote{At $230\,{\rm GHz}$, the characteristic EHT angular resolution of order $20\,\mu{\rm as}$ corresponds to about $4$--$5\,r_g$ for M87* and Sgr A*.} A resolved or image-averaged polarization fraction measures how polarized the image is locally, while an image-integrated or net fraction is often much smaller because polarized emission with different EVPAs or Stokes-$V$ signs cancels in the summed Stokes fluxes.

The electric vector position angle (EVPA), denoted $\chi$, is defined as
\begin{align}
    \label{eqn:evpadefn}
    \chi \equiv \frac{1}{2}\arg(P_\nu) = \frac{1}{2}\arg(Q_\nu + i U_\nu),
\end{align}
and reports the orientation of the electric field oscillation relative to the chosen sky basis, which we take to be east of north on the sky.\footnote{IAU; in standard astronomical images with north up and east left, this appears as rotations counterclockwise relative to vertical. See \citep{hamaker_1996_iaustokes} for more detail.} Linear polarization has no arrowhead, since rotating the electric field by $\pi$ gives the same physical state, so the EVPA is defined modulo $\pi$. The factor of one-half in Equation~\ref{eqn:evpadefn} thus converts the orientation of the quadratic Stokes pair $(Q_\nu,U_\nu)$ into an angle on the sky.

The observed Stokes vector is determined by solving for polarized radiative transfer through the plasma. For millimeter images of low-luminosity black hole accretion flows, the relevant radiative processes are synchrotron emission, absorption, and Faraday propagation effects \citep{rybicki_1979_radprocess,shcherbakov_2008_transfercoeffs,shcherbakov_2011_polcoeffs,leung_2011_transfercoeffs}. Neglecting the usually subdominant local scattering effects, the transfer equation along a ray can be written compactly for the invariant Stokes vector as
\begin{align}
    \label{eqn:polradxfer}
    \dfrac{d}{ds} \boldsymbol{\mathcal{I}}
    =
    \boldsymbol{\mathcal{J}}
    -
    \boldsymbol{\mathcal{K}}\boldsymbol{\mathcal{I}},
    \qquad
    \boldsymbol{\mathcal{I}}
    \equiv
    \dfrac{1}{\nu^3}
    \begin{pmatrix}
    I_\nu \\ Q_\nu \\ U_\nu \\ V_\nu
    \end{pmatrix},
\end{align}
where $s$ is an affine parameter, $\boldsymbol{\mathcal{J}}$ is the invariant emissivity vector, and $\boldsymbol{\mathcal{K}}$ is the invariant propagation matrix. The local plasma-frame coefficients that enter these objects are the usual emissivities $j_\nu$, absorptivities $\alpha_\nu$, and rotativities $\rho_\nu$, specified by the local magnetic field and lepton content, with invariant combinations that scale as $j_\nu/\nu^2$, $\nu\alpha_\nu$, and $\nu\rho_\nu$, respectively, with the factors of frequency $\nu$ accounting for gravitational and Doppler redshifts along the ray.

This form has a simple geometric interpretation. The source term $\boldsymbol{\mathcal{J}}$ adds a locally emitted Stokes vector along the ray, while the propagation matrix $\boldsymbol{\mathcal{K}}$ rotates and rescales the Stokes vector already present. In the usual local Stokes basis, the absorptivities enter the symmetric part of $\boldsymbol{\mathcal{K}}$, which attenuates and can mix polarization states: $\alpha_I$ sets the scalar optical depth, while the polarized absorptivities $\alpha_Q$, $\alpha_U$, and $\alpha_V$ describe dichroic absorption, which preferentially absorbs different polarization states and couples $I_\nu$ to $Q_\nu$, $U_\nu$, and $V_\nu$. The rotativities enter the antisymmetric part of $\boldsymbol{\mathcal{K}}$, which rotates the Stokes vector in polarization space: $\rho_V$ gives Faraday rotation between $Q_\nu$ and $U_\nu$, while $\rho_Q$ and $\rho_U$ give Faraday conversion between linear and circular polarization. Optical depth and Faraday propagation can therefore change both the polarized fraction and the relationship between the EVPA and the direction of the underlying magnetic field.

The local coefficients that populate $\boldsymbol{\mathcal{J}}$ and $\boldsymbol{\mathcal{K}}$ are microphysical inputs to the invariant transfer equation. For a specified magnetic field and lepton distribution functions, the synchrotron emissivities and absorptivities are computed by integrating the single-particle synchrotron emission and absorption kernels over the emitting charged-particle distribution. The Faraday rotation and conversion coefficients are likewise determined by the lepton distribution functions and charge composition, but through the dispersive plasma response rather than by summing emitted power directly \citep{rybicki_1979_radprocess}. Because direct kernel integrations are expensive, practical ray-tracing calculations do not typically perform them at every step and instead rely on analytic approximations, precomputed tables, or numerical fitting formul\ae{} for the coefficients before solving the polarized transfer equation along geodesics \citep{shcherbakov_2008_transfercoeffs,shcherbakov_2011_polcoeffs,leung_2011_transfercoeffs,dexter_2016_grtrans,pandya_2016_polcoeffs,marszewski_2021_transfercoeff,prather_2023_polcomp}.

Faraday effects are important because they both complicate interpretation of the EVPA and provide additional information about the source. One modeling simplification places the plasma that rotates the EVPA outside the emitting region in an \emph{external} screen so that the rotation follows a simple $\lambda^2$ dependence and can be treated as a correction to an intrinsic image. Under this assumption, derotating the EVPA recovers a source polarization pattern that can then be easily compared directly with models for the field geometry. Near a black hole, however, the electrons responsible for Faraday rotation and conversion may be mixed with the emitting region, spatially structured, relativistic, and variable, in which case the rotation measure (RM) is an \emph{internal} part of the source model rather than a foreground correction, and the same plasma conditions that determine the emitted polarization also determine how the polarization is rotated, converted, or depolarized before it escapes.

Circular polarization is similarly model dependent. It can be produced directly by intrinsic synchrotron emission or indirectly by Faraday conversion from linear polarization, although in the low-luminosity accretion flows observed by the EHT, Faraday conversion is generally expected to dominate at $230\,{\rm GHz}$. Because the relative importance of these two production channels depends on the plasma distribution, geometry of the magnetic field, and assumptions in the radiative transfer \citep{ricarte_2021_circularpol,eht_m87_9,joshi_2024_circularpol}, the sign and structure of Stokes $V$ carry model-dependent information about magnetic polarity and source orientation. But since Stokes $V$ is often weak and sensitive to calibration and to these modeling assumptions, it is difficult to use as a standalone diagnostic \citep{eht_m87_9,wielgus_2024_internalfaraday,joshi_2024_circularpol}.

\vspace{0.5em}

\begin{figure}[H]
%\begin{adjustwidth}{-\extralength}{0cm}
\centering
\includegraphics[width=1.\textwidth]{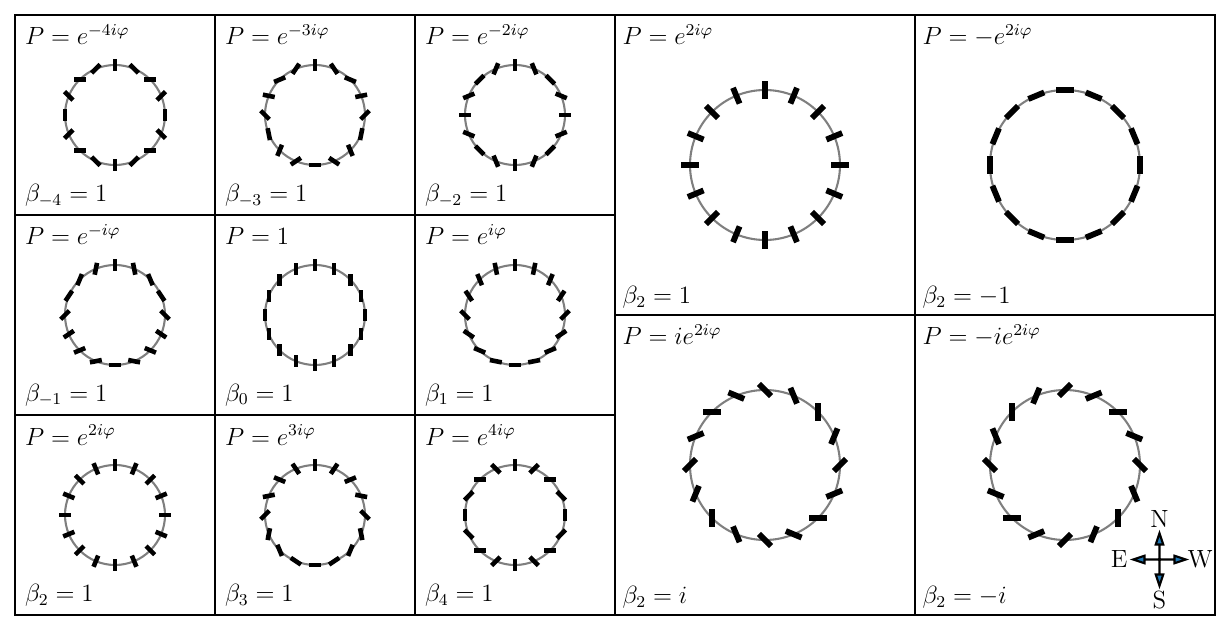}
%\end{adjustwidth}
\caption{Annular polarization modes used in the $\beta_m$ decomposition. The left panel shows pure azimuthal modes of $P=Q+iU$, and the right panel shows the special case $m=2$ for several phases of $\beta_2$. Radial, azimuthal, and handed spiral EVPA patterns are distinguished mainly by $\arg(\beta_2)$ for a nearly face-on ring. Figure reproduced with permission from \citep{palumbo_2020_beta2}.}
\label{fig:betam_schematic}
\end{figure}

The $\beta_m$ decomposition introduced by Palumbo, Wong, and Prather \cite[PWP;][]{palumbo_2020_beta2} provides a useful way to summarize the linearly polarized image structure. Writing the complex linear polarization as $P(\rho,\varphi)=Q(\rho,\varphi)+iU(\rho,\varphi)$ in polar image coordinates, the coefficients are annular Fourier moments,
\begin{align}
    \beta_m &\equiv \frac{1}{I_{\rm ann}}
    \int_{\rho_{\rm min}}^{\rho_{\rm max}}\int_0^{2\pi}
    P(\rho,\varphi)e^{-im\varphi}\rho\,d\varphi\,d\rho,\\
    I_{\rm ann} &\equiv
    \int_{\rho_{\rm min}}^{\rho_{\rm max}}\int_0^{2\pi}
    I(\rho,\varphi)\rho\,d\varphi\,d\rho .
\end{align}
The magnitude $|\beta_m|$ measures how much coherent polarized flux is in the $m$th azimuthal mode, while $\arg(\beta_m)$ gives the average rotation of the EVPA pattern relative to the chosen image axes, with the EVPA angle changing by half the corresponding phase change in $Q+iU$. Figure~\ref{fig:betam_schematic} presents a basic schematic of some of the pure $\beta_m$ modes. For nearly face-on, ring-like images, $\beta_2$ is the natural coefficient for the radial, azimuthal, or spiral component of the EVPA field because $Q+iU$ winds twice around the complex plane as a radial or azimuthal EVPA pattern winds once around the ring. In the narrow-ring limit, it is closely related to an $E/B$ decomposition of the linear-polarization pattern: real $\beta_2$ corresponds to an $E$-type radial or azimuthal pattern, while imaginary $\beta_2$ corresponds to a $B$-type handed spiral, up to the adopted sign convention and assuming absolute phase information \citep{palumbo_2020_beta2,eht_m87_7,eht_m87_8}. The exact values of the $\beta_m$ coefficients depend on the chosen annulus, image center, and coordinate orientation. If the source is not close to a face-on ring, if the polarized emission is strongly time variable, or if the relevant structure is not well captured by the chosen annulus, power can move between modes and $\beta_2$ no longer has the simple interpretation of a single coherent EVPA spiral \citep{narayan_2021_polring,gelles_2021_polmidplane,emami_2023_twisty,eht_sgra_8,palumbo_2025_spinconstraints,wong_2026_bhp2}.

The $\beta_m$ coefficients are image-domain summaries rather than raw interferometric observables. At the data level, the EHT measures complex correlations between pairs of telescopes, each of which samples a Fourier component of the sky brightness (after instrumental and atmospheric calibration). Polarimetric VLBI also measures correlations between receiver feeds, which may be circularly or linearly polarized depending on the station. Calibration procedures put these heterogeneous measurements into a common polarization basis, so that mixed-feed data can be combined before Stokes visibilities, fractional polarization, polarimetric ratios, or closure quantities are formed \citep[][for more detail, see also \citealt{martividal_2016_polconvert}]{eht_m87_3,eht_m87_7,ricarte_2023_resolvedpol}. Visibility-domain quantities are therefore closest to the measurement. Complex polarized visibilities preserve amplitude and phase information in the measured correlations, while fractional polarization, polarimetric ratios, closure quantities, and visibility-domain geometric model parameters can reduce sensitivity to station-based gain, phase, and leakage errors. For spin inference, a visibility-domain formulation is useful because it can be tested more directly against the data and against synthetic observations. For nearly face-on, ring-like sources, the same coherent EVPA structure summarized by image-domain $\beta_2$ can also be tested directly in the visibility domain through polarimetric phases \citep{palumbo_2023_ringpolarimetry}, providing a direct bridge between compact image descriptors and the VLBI data.

Turning the underlying visibility data into an image is a complicated, underspecified problem, however. The image reconstruction procedure combines sparse, noisy, and calibration-dependent Fourier-domain data with assumptions about smoothness, compactness, positivity, polarimetric structure, and time variability, and the reconstruction procedures must be tested against synthetic data and alternative algorithms \citep{eht_m87_7,eht_sgra_7}. Image-domain summaries should therefore be thought of as compact descriptions of data-constrained structure rather than as primitive observables. They are valuable because they organize complicated polarization structure, but a proposed spin-sensitive feature may be stronger when it can also be checked directly in the visibility domain or tested with synthetic observations, as in long-baseline polarimetric phase diagnostics for the photon ring \citep{palumbo_2023_ringpolarimetry}.

\vspace{0.5em}

Polarimetry therefore gives the most physical insight into the near-horizon plasma structure rather than spin itself. The conceptual pipeline is as follows: the data are correlations and visibilities, from which we reconstruct or fit images, ring parameters, polarization fractions, EVPA patterns, circular polarization structure, and compact summary statistics like $\beta_2$. Those quantities constrain the structure of the magnetic field, Faraday depth, emission geometry, accretion state, and plasma conditions. The spin inference comes later, by asking which of those constraints can be connected to $\bhspin$ given assumptions about the source and the radiative transfer. This measurement hierarchy is the organizing thread for the rest of the article: a spin-sensitive mechanism is useful when it can be tied to a measured or reconstructed polarimetric quantity and the source assumptions can be explicitly specified.

\section{Connections between spin and polarization}
\label{sec:connection}

At radio frequencies, emission near the black hole event horizon is produced by synchrotron radiation from electrons (and possibly positrons) in magnetized plasma near the black hole. The magnetic field sets the orientation of the locally emitted polarization, while relativistic motion and radiative-transfer effects modify the signal near the source. Propagation through curved spacetime, including parallel transport of the polarization basis, then determines how that signal appears to the observer in the Stokes images \citep{connors_1980_xraypol}. Spin affects the observed polarimetric signature via its influence both on the propagation of polarized light and on the magnetic field carried by the accreting plasma. Here we focus on which parts of the measured polarization can be traced to spin-sensitive field structure and which are controlled by plasma physics, propagation, variability, or observational uncertainty. We consider this question in different observational regimes and at different characteristic scales: resolved EHT emission-region polarimetry on scales of a few $r_g$, jet-base polarimetry and the disk--jet connection on scales of tens to hundreds of $r_g$, near-horizon EVPA behavior, and photon-ring or subring signatures accessible through long-baseline polarimetry.

\subsection{Horizon-scale images: resolved polarization on scales of a few $r_g$}

Resolved horizon-scale polarimetry constrains the magnetic field structure within a few gravitational radii of the black hole, but that field cannot be uniquely determined by the image features. In flat space and in the optically thin limit, the observed EVPA would be perpendicular to the projected magnetic field. In an actual 230 GHz image, however, the observed EVPA is also affected by relativistic motion, lensing, parallel transport, absorption, Faraday rotation, and stochastic variability in time arising from turbulence. Circular polarization and Faraday conversion provide additional information about magnetic polarity, line-of-sight structure, electron thermodynamics, and optical depth, but they are also more sensitive to transfer and calibration systematics. A fully polarized image is therefore not a direct map of the underlying magnetic field and instead records how an emitting, magnetized plasma in motion appears after polarized radiative transfer through curved spacetime.

The inferred structure of the magnetic field is astrophysically interesting because it constrains the plasma and accretion state. The EHT polarimetric images of M87* show an ordered ring-scale EVPA pattern, and comparisons with numerical simulations typically prefer models with a strong magnetic flux. The most consistent models are usually in the magnetically arrested disk (MAD) accretion state \citep{bisnovatyi_1974_madstar,igumenshchev_2003_mad,narayan_2003_mad}, since they best satisfy constraints from both the polarimetric morphology and the overall jet power \citep{eht_m87_7,eht_m87_8}. The EHT polarimetric images of Sgr A* also show a bright polarized ring and a coherent spiral EVPA pattern, but the source evolves during an observing night and the degree to which Faraday rotation affects the source is known to change with time
\citep{eht_sgra_7,eht_sgra_8,goddi_2021_ehtpoltargets,bower_2018_almapolsgra,
marrone_2007_faraday,wielgus_2022_sgrapolalma,wielgus_2024_internalfaraday}. Multi-epoch M87* polarimetry suggests a similar caution on longer timescales: the observed polarized morphology changes, so candidate spin-sensitive structures should be interpreted as properties of a magnetized, variable accretion state rather than fixed labels of the spacetime \citep{eht_m87_20172021}. Circular polarization provides related constraints on magnetic polarity and Faraday conversion, but current detections and upper limits remain difficult to interpret without a detailed, robust model for the radiative transfer \citep{eht_m87_9,joshi_2024_circularpol,ricarte_2021_circularpol,munoz_2012_sgracircularpol,bower_2018_almapolsgra}.

The spin dependence of these observables is the result of a causal chain rather than something that can be read directly from the image. In MADs, the magnetic field is strong enough to affect the dynamics of the inflow, and the field geometry is shaped by the combination of poloidal flux accumulated on the horizon and toroidal field produced and supported by the disk \citep{narayan_2003_mad,tchekhovskoy_2011_mad,mckinney_2012_mad,begelman_2022_whatmad}. Regularity of the magnetic field at the event horizon, frame dragging, and the rotating inflow determine the $B^\phi/B^r$ pitch of the magnetic field in the emitting region, which then shapes EVPA spirals and compact descriptors such as $\arg(\beta_2)$ \citep{palumbo_2020_beta2,emami_2023_twisty,chael_2023_bhp1,hou_2025_nearhorizonpol,wong_2026_bhp2}. In simple, nearly face-on, optically thin cases with controlled transfer, a spiral EVPA pattern is therefore sensitive to the ratio between toroidal and radial magnetic-field components, $B^\phi/B^r$; in realistic images it is a model-dependent tracer of that ratio rather than a direct measurement. The PWP coefficient $\beta_2$ was introduced to quantify coherent, ring-scale polarization structure, and that paper showed that large $|\beta_2|$ can distinguish highly magnetized MAD image morphologies from weaker-field standard and normal evolution (SANE) models, exhibiting systematic behavior across a library of numerical simulations.

Wong et al. \citep[BHP II;][]{wong_2026_bhp2} made a link in this spin--field--polarization chain more explicit. In the force-free limit, horizon regularity and the rotation of the black hole constrain the ratio between the radial and toroidal components of the magnetic field, $B^\phi/B^r$, on the event horizon, and for a face-on viewing geometry, the phase $\arg(\beta_2)$ is sensitive to essentially the same ratio \citep[BHP I;][]{chael_2023_bhp1}. Away from the horizon, however, the field structure is determined by both this boundary condition and the dynamics of the plasma, with numerical models predicting a significant toroidal field component \citep{begelman_2022_whatmad,zhang_2024_mads}. By self-consistently solving for stationary semianalytic inflow models, BHP II showed how finite inertia modifies this relation: as the accreting plasma becomes less magnetically dominated, the flow drags and winds the field, shifting the polarization pattern away from the ideal force-free expectation in a predictable way. The competition between horizon-imposed field winding and advection by the rotating inflow gives a physical explanation for trends seen in numerical image libraries. The spin parameter sets the relevant near-horizon boundary conditions, the magnetized inflow supplies the plasma dynamics, and polarimetry is sensitive to the resulting magnetic pitch.

Yet in realistic source models, this clean connection between the magnetic field and the black hole spin is complicated by the many other system parameters. Numerical simulations have therefore been used to generate large image libraries that vary spin, magnetic flux state, inclination, field polarity, electron thermodynamics, and radiative transfer together; see Figure~\ref{fig:survey_spin_gallery} for an example of such a library. Quantitative analysis of these libraries and comparison against observations have often been used to inform constraints on the magnetic flux state, Faraday depth, and inclination much more robustly than $\bhspin$ itself \citep{chael_2019_twotemp,dexter_2020_sgrasurvey,mizuno_2021_twotemp,wong_2022_patoka,qiu_2023_mlpol,chatterjee_2023_ngeht_radiation,zhang_2024_mads,chael_2025_radsurvey,dhruv_2025_surveyv3,eht_m87_5,eht_m87_8,eht_sgra_5,eht_sgra_8,eht_m87_2018_2}.

More simplified geometric source models play a complementary role by isolating the ingredients responsible for particular polarimetric signatures. Rings, equatorial emitters, and hotspots can specify the source and field geometry precisely enough to show how orbital motion, lensing, parallel transport, and time dependence generate EVPA spirals or periodic ``looping'' behavior in the $Q$--$U$ plane \citep{broderick_2005_hotspots,gelles_2021_polmidplane,gravity_2018_orbitisco,wielgus_2022_sgrapolalma,vos_2022_polhotspot,vincent_2024_polhotspot,yfantis_2024_quloops,ricarte_2025_sgrapolcurves}. Their value is therefore not that they provide model-independent spin observables, but that they make the required assumptions explicit. A loop, pitch angle, or handedness is not a spin measurement until the emitting structure, velocity field, and magnetic field geometry have been specified. The same caveat applies to attempts to associate horizon-scale millimeter polarimetry with a particular orbital radius, such as the innermost stable circular orbit (ISCO). The ISCO is a geodesic landmark and an effective inner edge in idealized thin-disk models, but pressure-supported tori and magnetized accretion flows can have cusps, stresses, and radiative or emissive edges that do not coincide with it \citep{fishbone_1976_torus,kozlowski_1978_donut,gammie_1999_inflow,krolik_2002_inneredge,beckwith_2008_radiationedge}. Especially in the hot accretion disks observed by the EHT, the observed emission need not truncate at the ISCO, and the polarized morphology is shaped by the emitting plasma, magnetic field, transfer effects, and lensing rather than by a single privileged radius \citep{narayan_1995_ADAF,yuan_2014_review,wong_2025_mixing,chael_2021_innershadow,eht_m87_5,eht_sgra_5,palumbo_2025_spinconstraints}. In this sense, reduced models that connect the EVPA pitch to the direct and indirect image orders are complementary to BHP-style calculations: the former expose which source assumptions are needed to interpret a given polarimetric pattern while the latter attempt to derive the magnetic pitch from horizon regularity and the dynamics of the magnetized inflow \citep{narayan_2021_polring,gelles_2021_polmidplane,vos_2022_polhotspot,vincent_2024_polhotspot,ricarte_2023_resolvedpol,palumbo_2025_spinconstraints}.

\vspace{0.5em}

These spin-sensitive image summaries are also modified by non-stationary plasma dynamics and Faraday effects \citep{shcherbakov_2012_sgrapol,moscibrodzka_2017_m87faraday,jimenezrosales_2018_faraday,wielgus_2024_internalfaraday}. Internal Faraday rotation can shift $\arg(\beta_2)$, reduce $|\beta_2|$, lower the resolved polarized fraction, and make the apparent RM frequency dependent, so that the EVPA cannot be corrected by a single foreground-screen model. The relevant Faraday depth is set by the column of Faraday-active plasma, the magnetic field along the line of sight, and the electron temperature, so changes in accretion rate or thermodynamics can rotate or depolarize the signal even if the underlying field geometry is unchanged. Faraday conversion can generate circular polarization from linear polarization, and because conversion depends on the orientation and polarity of the magnetic field along the ray, the supply of low-temperature Faraday-active electrons, and optical depth, Stokes $V$ can help constrain magnetic polarity and plasma content \citep{ricarte_2021_circularpol,joshi_2024_circularpol,tsunetoe_2020_m87pol,eht_m87_9}. For Sgr A*, the variable RM and image-polarization analysis suggest that much of the rotation may be produced within or close to the emitting region rather than by a passive foreground screen, so it is part of the source model rather than a removable foreground correction \citep{wielgus_2024_internalfaraday}. For M87*, the importance of Faraday effects is less constrained, and circular-polarization detections and changes to the EVPA on intermediate timescales show that propagation and variability remain an unavoidable part of the inference modeling problem \citep{eht_m87_9,eht_m87_20172021}.

\subsection{The jet base and disk--jet connection: tens to hundreds of $r_g$}

The same spin--field connection can also be followed out of the compact emission region and into the jet. Jet base polarimetry is less direct as a probe of the near-horizon spacetime, but the black-hole-related polarimetric signatures can extend over tens to hundreds of gravitational radii, where they may be easier to resolve. The connection between the compact emission region and the emerging jet is itself model dependent, since the observed jet-base polarization may sample emission associated with horizon-threading flux, disk-launched material, a jet sheath, or some mixture of these components. In stationary, axisymmetric ideal magnetohydrodynamics or force-free outflow models, magnetic flux surfaces rotate with a field-line angular velocity $\Omega_F$. The associated \emph{light cylinder radius} (or more generally, the outer light surface in curved spacetime) marks where a point corotating with the flux surface would move at the speed of light. In the flat, cylindrical limit, this scale is $R_{\rm LC}\sim c/|\Omega_F|$, and outside of it, rotation winds field lines so that the toroidal magnetic-field component typically becomes important. Near this scale, the projected magnetic-field structure and therefore the EVPA can change rapidly \citep{blandford_1977_bz,znajek_1977_condition,gralla_2014_ffmagnetospheres,narayan_2022_jetsurvey,gelles_2025_spinjetpol,gelles_2026_offaxisjetpolspin}. In idealized BZ-like magnetospheres, $\Omega_F$ is tied to the horizon angular velocity, $\Omega_H=\bhspin/[2(1+\sqrt{1-\bhspin^2})]$ here expressed in units of $t_g^{-1}$
where $t_g\equiv r_g/c$, and is often of order $\Omega_H/2$ \citep{tchekhovskoy_2011_mad,mckinney_2012_mad}. If this relation holds, then the location of the polarimetric transition can carry information about $\bhspin$ \citep{gelles_2025_spinjetpol,gelles_2026_offaxisjetpolspin}.

\begin{figure}[H]
\centering
\includegraphics[width=1.\textwidth]{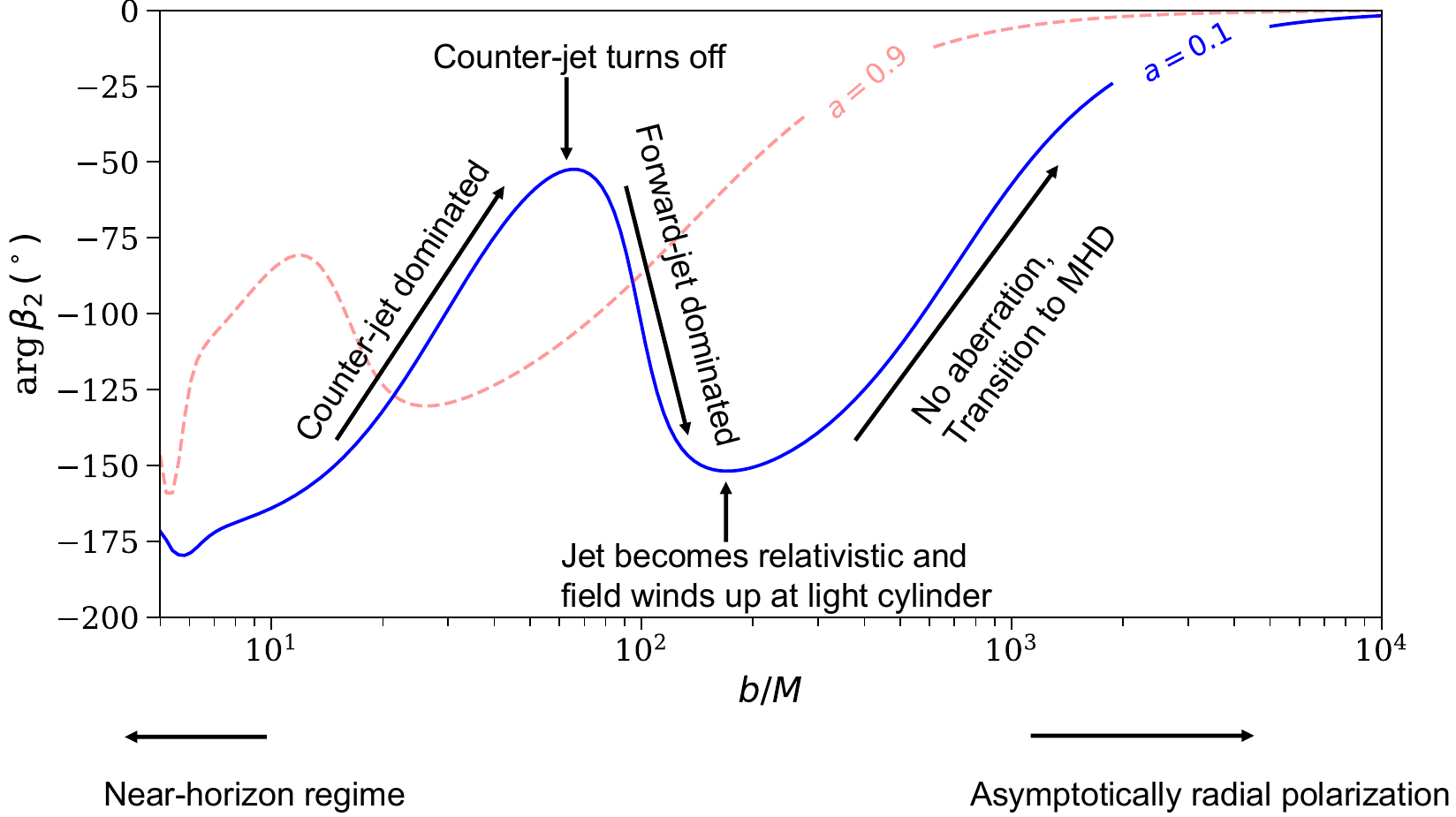}
\caption{Radial evolution of EVPA morphology in nearly face-on jet models. The curves show $\arg(\beta_2)$ measured as a function of impact parameter $b$ in units of $r_g$ for low- and high-spin cases. Large swings occur where the counterjet fades, where the magnetic field becomes azimuthally dominated near the light cylinder, and where the flow approaches its terminal Lorentz factor; the light-cylinder swing is the feature most directly tied to the field-line rotation and hence to spin. Figure reproduced with permission from \citep{gelles_2025_spinjetpol}.}
\label{fig:jetevpa}
\end{figure}

Gelles et al.~developed this idea into an explicit polarimetric observable for nearly face-on jets, including M87*, where the jet is understood to be viewed at small inclination \citep{mertens_2016_m87jet,walker_2018_m87jet,hada_2024_m87review}. In their models, large spatial swings in the EVPA occur at several characteristic radii: where the counterjet fades, where the magnetic field becomes azimuthally dominated near the light cylinder, and where the plasma reaches its asymptotic Lorentz factor \citep{gelles_2025_spinjetpol}. The light-cylinder feature is the one most directly connected to spin because its position is controlled by $\Omega_F$, which is in turn tied to the black hole spin; see Figure~\ref{fig:jetevpa}. For more inclined jets, the same transition appears through asymmetric spine and limb polarization. The predicted swing depends on the jet geometry, acceleration profile, emitting plasma, optical depth, Faraday effects, and viewing angle \citep{gelles_2026_offaxisjetpolspin}.

\vspace{0.5em}

Even when the light-cylinder transition is not resolved, jet polarimetry can still provide partial information about the sign of the black hole spin or the magnetic topology. In a jet threaded by a helical magnetic field, transverse EVPA structure, rotation-measure gradients, and the sign of circular polarization can all indicate how the projected field winds on the sky \citep{asada_2002_helical_3c273,lisakov_2021_3c273,casadio_2019_cta102,goddi_2025_m87jetalma,park_2025_m87helix}. These observables do not give an absolute spin direction by themselves, but they can constrain this magnetic handedness. Stokes $V$ and the sign of Faraday rotation provide complementary information because both are sensitive to the polarity and orientation of the magnetic field along the line of sight in the emitting or Faraday-rotating plasma \citep{gabuzda_2018_bfieldrotation,ricarte_2021_circularpol}. To turn field handedness into a spin direction, one must also infer which side of the jet is approaching, how the disk and jet are oriented on the sky, and whether the observed emission samples field lines wound by the central engine rather than by a sheath, shock, external screen, or disk-launched outflow. M87* is a useful test case for this procedure, since horizon-scale polarimetry, jet power, and jet polarization on larger scales can in principle be oriented within the same, global magnetic field model. Yet the diagnostic remains model dependent: magnetic polarity, sheath emission, external Faraday rotation, shocks, instabilities, and disk-launched material can all alter the observed handedness without changing the black hole spin \citep{hada_2024_m87review,kino_2022_m87slowmag,park_2025_m87helix,goddi_2025_m87jetalma}.

\vspace{0.5em}

Taken together, jet measurements thus provide another test of the connection between spin and the magnetic field. The case is stronger when near-horizon EVPA morphology, circular polarization, Faraday depth, jet power, and a light-cylinder or helical-field signature in the resolved jet support a consistent physical picture across these scales, rather than resting on any one observable alone. Horizon-resolved sources like M87* and Sgr A* can therefore play a calibrating role by connecting spin-sensitive polarimetry near the black hole to jet signatures that may be measurable in a larger sample. Conversely, a mismatch between the ring and jet polarization would be astrophysically informative even if it weakens a direct spin inference, because it would identify where plasma loading, dissipation, or jet collimation breaks the simple connection between the horizon-threading field and the observed radiation.

\subsection{Near-horizon EVPA structure and the inner shadow}

\begin{figure}[H]
\centering
\includegraphics[width=1.\textwidth]{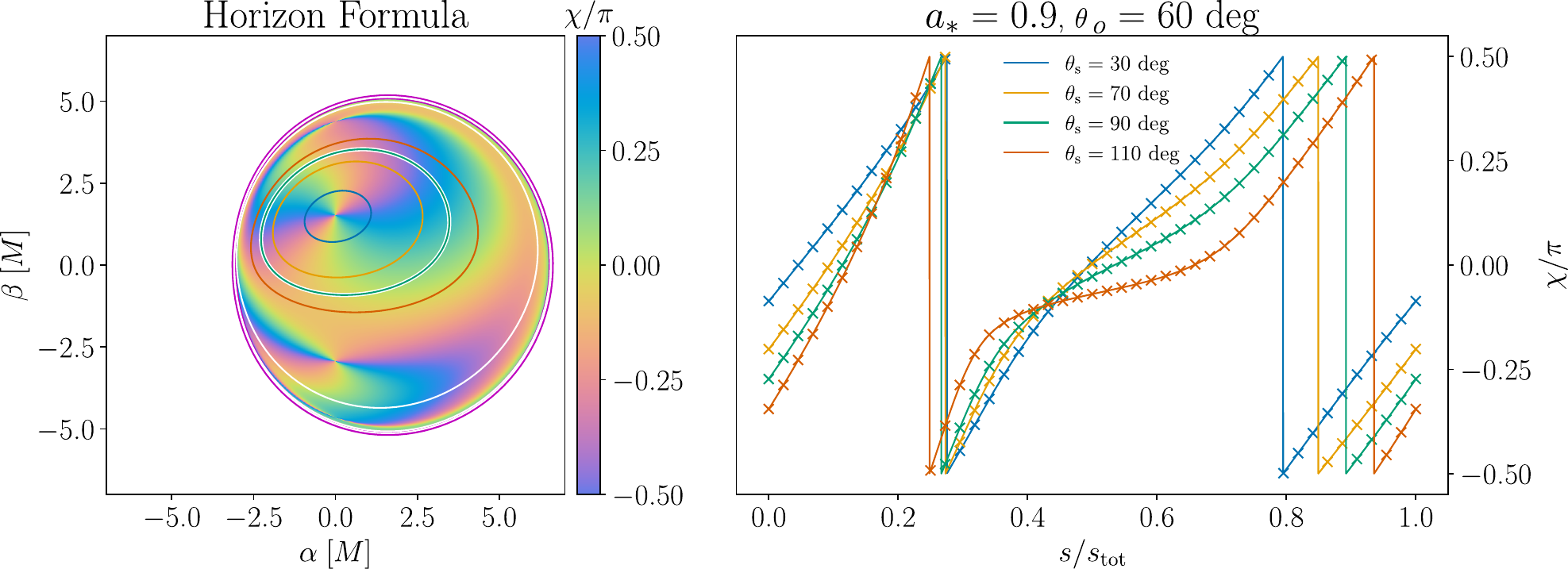}
\caption{Near-horizon EVPA pattern predicted by the horizon-limit formula. The color maps show $\chi/\pi$ on the observer screen for a rapidly spinning black hole viewed at $\theta_o=60^\circ$, with curves marking image tracks for several emission latitudes. Reproduced with permission from \citep{chael_2026_bhp3}.}
\label{fig:bhp3_signature}
\end{figure}

The most compact parts of the image offer a complementary probe of the connection between spin and the magnetic field. Polarization from direct emission still depends on uncertain emissivity, turbulence, and flow geometry, but close to the event horizon the observed EVPA becomes more strongly controlled by near-horizon magnetic-field structure and by the projection of that structure through curved spacetime onto the observer's sky. This limit is most relevant near the inner shadow, the central brightness depression associated with rays that fall directly into the event horizon without winding around the hole \citep{chael_2021_innershadow}. The resulting signature still requires a source model, but it emphasizes image regions where detailed plasma structure may be less dominant. BHP I first emphasized that polarization near the inner shadow can carry information about horizon-threading electromagnetic structure \citep{chael_2023_bhp1,chael_2021_innershadow}, and Hou et al.~later derived the direct-image, equatorial limit of this EVPA pattern for stationary, axisymmetric, degenerate equatorial synchrotron emitters \citep{hou_2025_nearhorizonpol}. In particular, they showed that the near-horizon EVPA is fixed by the spacetime and observer inclination rather than by detailed flow structure. \citet[BHP III;][]{chael_2026_bhp3} then extended this result to arbitrary horizon latitude and all image orders and showed how the limiting behavior arises physically because the magnetic field approaches a predominantly toroidal orientation near the horizon, which is mapped onto the image plane by parallel transport through the curved spacetime. Figure~\ref{fig:bhp3_signature} illustrates this spin-dependent limiting pattern on the horizon.

The observational difficulty with this polarimetric signature is that it is faint and easily contaminated. Emission from plasma arbitrarily close to the horizon is strongly redshifted, foreground emission can dominate the image inside the critical curve, and Faraday rotation or off-equatorial emission can move the measured EVPA away from the limiting value. The more practical target may therefore be the radial evolution of the polarization as the image approaches the inner shadow. In MAD simulations, especially when the emission is concentrated near the equatorial plane, the EVPA or $\arg(\beta_2)$ has been shown to swing toward the limiting horizon value at small impact parameter. BHP III argues that space-VLBI experiments such as the Black Hole Explorer \citep[BHEX;][]{johnson_2024_bhex} could detect this trend in M87* if the near-horizon emission is sufficiently equatorial and the Faraday propagation effects can be well modeled \citep{chael_2026_bhp3}.

\subsection{The photon ring, subrings, and long-baseline signatures}

Photon-ring subimages provide a complementary polarimetric probe because strong lensing ties their structure more directly to spacetime geometry than to the detailed dynamics of the emitting plasma. The full image can be decomposed into a sequence of subimages, where the direct image is labeled $n=0$ and is composed of all photons that propagate directly to the observer without passing around the far side of the black hole. Higher-order subimages, with $n\ge1$, are produced by photons that pass close to unstable bound photon trajectories and execute partial turns around the black hole before reaching the observer. These higher-order subimages, or subrings, accumulate near the critical curve, the limiting image-plane curve approached by photons that spend many orbits near unstable bound trajectories in the black hole spacetime \citep{teo_2003_photonskerr,gralla_2020_kerrlensing,johnson_2020_universal,lupsasca_2024_guidegr}. Their locations are therefore governed more directly by Kerr geometry than those of the direct image, and their polarization carries a self-similar structure set by parallel transport near the photon orbit \citep[][see also \citealt{okten_2025_spinodd} for a related analysis]{himwich_2020_universalpol}. Numerical simulations show that this structure is informative but not simply universal: in many models, the photon-ring region is partially depolarized by interference between differently oriented lensed polarization patterns, and the amount of depolarization depends on magnetic flux state, spin, and emission geometry \citep{palumbo_2022_photonringbeta2}.

\begin{figure}[H]
\centering
\includegraphics[width=1.\textwidth]{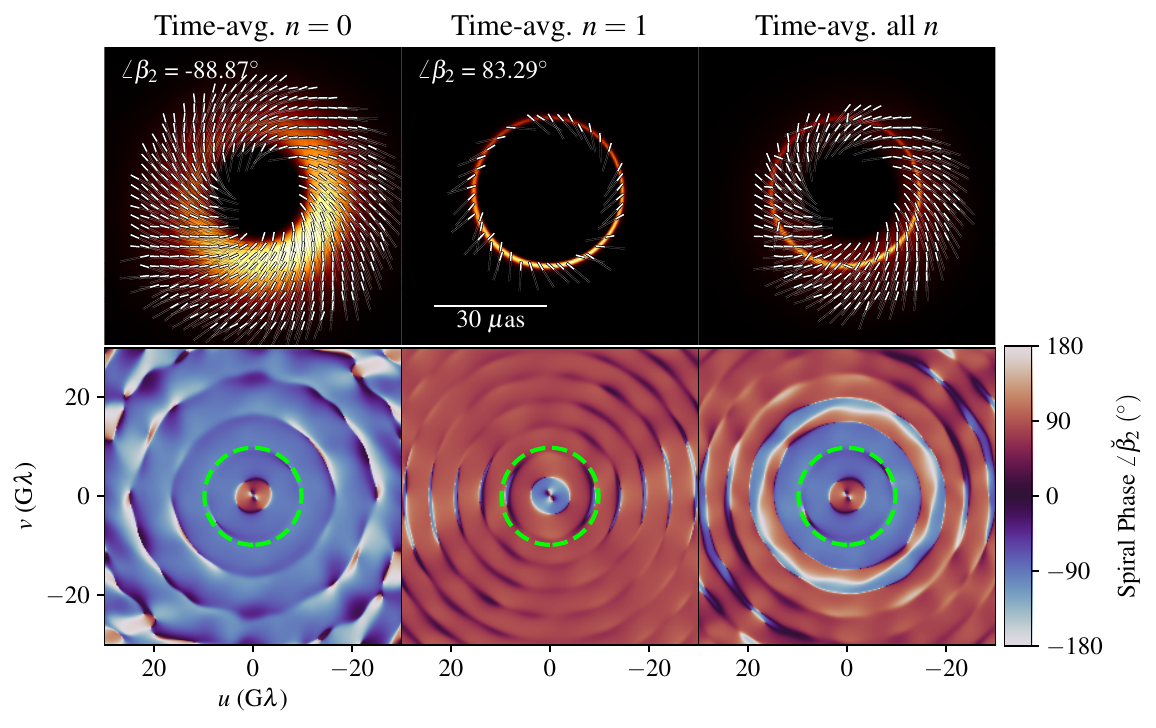}
\caption{Time-averaged decomposition of a simulated polarized image into the direct image ($n=0$), the first lensed subimage ($n=1$), and the full image, together with the corresponding visibility-domain spiral phase. The figure illustrates how long-baseline polarimetry can distinguish signal in the direct image from signal due to the photon subring contributions. Reproduced with permission from \citet{palumbo_2023_ringpolarimetry}.}
\label{fig:palumbo2023_tavg_decomp}
\end{figure}

Long-baseline polarimetric observables, accessible through space-VLBI or higher-frequency ground-based observations, are natural probes of the transition from the direct image to photon-ring structure. As illustrated in Figure~\ref{fig:palumbo2023_tavg_decomp}, the decomposition by subring image order also has a signature in the visibility domain: the transition from the direct image to the lensed subimages leads to changes in the polarimetric phase on long baselines. Baseline length therefore provides a useful proxy for approximate angular-scale separation. Shorter baselines primarily constrain the broader direct ($n=0$) image and its $\beta_2$-like EVPA structure with less contamination from compact lensed subimages, while longer baselines can target the indirect or higher-order lensed ($n\ge 1$) subimages near the critical curve. \citet{palumbo_2023_ringpolarimetry} proposed gain-robust polarimetric phase diagnostics for this transition, estimating that M87* requires roughly mJy sensitivity on baselines longer than $15\,{\rm G}\lambda$ at 230 GHz, while Sgr A* likely requires higher-frequency observations because refractive interstellar scattering obscures the long-baseline photon-ring signal at 230 GHz \citep{johnson_2016_scattering,narayan_1992_diffrefr,johnson_2016_refractive,palumbo_2023_ringpolarimetry}. 

A subsequent study of proposed next-generation Event Horizon Telescope (ngEHT) array improvements showed that repeated polarimetric observations of Sgr A* on the longest Earth-based baselines could detect the photon ring \citep{shavelle_2024_sgra_photonring}. A follow-on GRMHD library survey of 230 GHz images of M87* found a broader class of total-intensity, linear-polarization, and circular-polarization transitions in long-baseline observables that may enable photon-ring detection, though in a smaller fraction of accretion flows than expected for Sgr A* \citep{tamar_2024_photonringpol}; these transitions are expected to move to shorter baselines at higher frequencies. Photon-ring polarimetry therefore points naturally to space-VLBI missions such as BHEX and to higher-frequency ground arrays \citep{doeleman_2023_ngehtarray,johnson_2024_bhex,raymond_2024_eht345,ngeht_2025_fundamentalphysics}. Such low-dimensional visibility-domain observables may be especially useful for near-future population studies, where ngEHT-like data may provide spin-sensitive source-structure proxies for many systems rather than detailed image reconstructions for only M87* and Sgr A* \citep{pesce_2022_expectationsngeht}. Nevertheless, although such a detection would isolate the image orders whose geometry is most directly controlled by the Kerr photon shell, it would not by itself be a spin measurement. With sufficient sensitivity and a source-polarization model, however, the shape, displacement, width, and parallel-transported EVPA structure of these subimages could provide spin-sensitive constraints.

\section{Current constraints}
\label{sec:current}

\begin{figure}[H]
\centering
\includegraphics[width=0.8\textwidth]{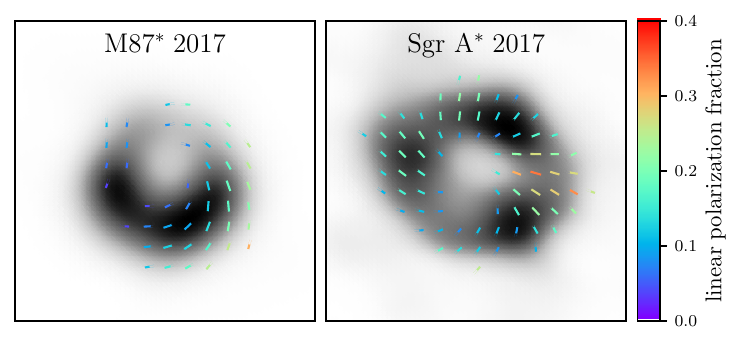}
\caption{Event Horizon Telescope polarimetric images of M87* and Sgr A* from the 2017 observations. The background grayscale maps show total intensity, line segments show EVPA, and segment color gives the resolved linear polarization fraction. Both sources show compact ring-like emission with organized linear polarization, but their polarization fractions, variability, and source-model interpretations differ \citep{eht_m87_7,eht_sgra_7}.}
\label{fig:eht_data}
\end{figure}

EHT images have made spin inference more concrete by enabling direct comparisons between data and theoretical models of the emitting plasma and magnetic field. Figure~\ref{fig:eht_data} shows representative 2017 EHT polarimetric images for M87* and Sgr A*. The measured ring diameters of the two sources fix the angular scale of the compact emission relative to independently measured black hole masses, while polarimetric measurements provide information about field order, Faraday depth, magnetic flux state, electron thermodynamics, and circular polarization \citep{eht_m87_1,eht_m87_5,eht_m87_7,eht_m87_8,eht_m87_9,eht_sgra_1,eht_sgra_5,eht_sgra_7,eht_sgra_8}. A useful way to organize current claims is therefore to separate what the data directly constrain, what source-model properties are inferred from those constraints, and which of those inferences can be translated into evidence about $\bhspin$.

\subsection{M87*}

The observational starting point for M87* is the 2017 EHT linearly polarized images, which present a resolved ring with a coherent, nearly azimuthal EVPA pattern and peak resolved fractional linear polarization $|m|$ of order $15\%$ \citep{eht_m87_7}. The EHT reports an image-integrated net linear polarization fraction $|m|_{\rm net}\simeq 1.0$--$3.7\%$, an image-averaged resolved linear polarization fraction $\langle |m| \rangle\simeq 5.7$--$10.7\%$, $|\beta_2|\simeq 0.04$--$0.07$, and $\arg(\beta_2)\simeq -163^\circ$ to $-127^\circ$ \citep{eht_m87_7}. The 2017 EHT data also constrain the image-averaged resolved circular polarization fraction to $\langle |v| \rangle < 3.7\%$, consistent with the simultaneous ALMA image-integrated circular polarization limit $|v_{\rm int}|<1\%$ on larger angular scales \citep{eht_m87_9}. Stokes $V$ is physically related to the direction of the magnetic field and Faraday conversion.

Interpreting these measurements requires a source model; when polarimetry is added to total intensity and ALMA constraints, the EHT model comparison favors MAD states with dynamically important magnetic fields, and within that sampled library, imposing the requirement that the simulated jet power be of the same order as the observationally inferred jet power leaves only strongly magnetized, nonzero-spin models \citep{eht_m87_5,eht_m87_8}. The 2017 EHT data therefore provide a conditional inference about source parameters that is tied to the simulated library and jet power cut.

The 2017, 2018, and 2021 EHT data provide additional information: the ring diameter is stable, but the total intensity morphology and linear polarization are not. The resolved fractional linear polarization peaks near $15\%$ in 2017 but near $5\%$ in 2018 and 2021. The EVPA geometry in 2018 differs substantially from both 2017 and 2021, whereas the difference between 2017 and 2021 could reflect intrinsic flow evolution or a changing Faraday screen \citep{eht_m87_20172021}. The 2018 total-intensity data still recover a persistent ring, with median diameter $43.3^{+1.5}_{-3.1}\,\mu{\rm as}$ and a brightness-asymmetry position angle shifted by about $30^\circ$ relative to 2017 \citep{eht_m87_2018_1}. That same epoch was quasi-contemporaneous with a gamma-ray flare in M87 \citep{eht_m87_20172021,hada_2024_m87review}. Together these epochs constrain magnetic structure and variability, and imply a possible multiwavelength coupling while ruling out an overly simple stationary-source interpretation; they motivate simultaneous gamma-ray, X-ray, millimeter, and horizon-scale polarimetric monitoring.

The first informative spin constraint comes in the form of a statement about its orientation: if the black hole spin is aligned or anti-aligned with the jet observed at large scales, EHT model comparisons infer that the M87* spin vector points away from Earth \citep{eht_m87_5,eht_m87_2018_2}. Comparing the 2017 and 2018 images adds more information to the interpretation in that the observed change in the location of the ring brightness asymmetry is naturally produced by turbulent accretion; within the sampled 2018 model library, the more turbulent retrograde models reproduce the multi-epoch brightness-asymmetry shift better than the prograde models, while the tilted models in that library are inconsistent with the data \citep{eht_m87_5,eht_m87_2018_2}. A separate analysis of ring asymmetry provides a weak constraint on the magnitude, marginally disfavoring low-spin MAD models with $|\bhspin|\lesssim 0.2$, but this inference is conditioned on the magnetic flux state of the accretion, turbulent realizations, flow orientation, and the statistic used to quantify the image asymmetry \citep{bernshteyn_2026_asymmetry}.

\begin{figure}[H]
\centering
\includegraphics[width=\textwidth]{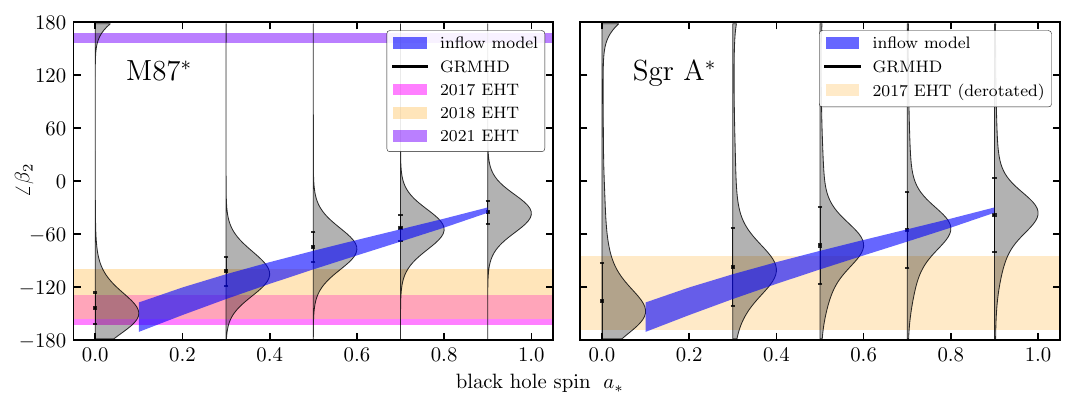}
\caption{Comparison between EHT constraints on $\arg(\beta_2)$ and semianalytic model predictions from \citet{wong_2026_bhp2}. The purple bands show the BHP II stationary inflow model as a function of prograde spin, the gray distributions show GRMHD image results at discrete spins, and the horizontal bands show EHT measurements for M87* and derotated Sgr A*. Within this model family and its assumptions about emission geometry and Faraday rotation, the observed bands lie away from the high-spin prograde predictions. Figure reproduced with permission from \citep{wong_2026_bhp2}.}
\label{fig:bhp2_comparison_both}
\end{figure}

Considering the EVPA morphology provides more spin-magnitude-sensitive information. BHP II, which restricts its model space to prograde accretion flows, relates $\arg(\beta_2)$ to the pitch of the magnetic field, $B^\phi/B^r$, and disfavors high-spin configurations for M87* under assumptions about Faraday rotation and emission geometry \citep{wong_2026_bhp2}, although those models do not include retrograde configurations. Figure~\ref{fig:bhp2_comparison_both} shows this comparison between observed $\arg(\beta_2)$ ranges and model predictions. The radiative, two-temperature survey of MADs performed by Chael suggests a similar preference for retrograde or only modestly prograde spins, although its images overproduce linear polarization \citep{chael_2025_radsurvey}. A random-forest analysis of resolved polarimetric observables found that, among model images passing M87* constraints, most were retrograde and the posterior favored high-spin retrograde models with large ion-to-electron temperature ratios \citep{qiu_2023_mlpol}. Another machine-learning-based analysis of 2017 EHT synthetic GRMHD visibilities similarly found M87* best described by a retrograde MAD configuration with $|\bhspin|$ in the approximate range $0.5$--$0.94$ \citep{janssen_2025_zingularity3}. The spread among these studies identifies the dominant systematics: Faraday effects, electron thermodynamics, emission geometry, and the training or simulation library all influence how a measured EVPA pattern is translated into $\bhspin$. M87* therefore illustrates both the progress and the present limit of polarimetric spin inference: the data now constrain the magnetic state strongly, but converting that constraint into a spin magnitude still depends on modeling assumptions.

Part of this spread may reflect physical differences between model families rather than statistical or library-level uncertainty. The observed polarization is shaped both by the magnetic field at emission and by transfer through the surrounding plasma, so different prescriptions for field geometry, inflow, and Faraday-active material can move the inferred spin even when the same image features are being compared. Polarized-image models have shown that magnetized plasma near the hole can rotate and depolarize linear polarization, while Faraday conversion can generate circular polarization whose sign and morphology depend on the geometry of the magnetic field along the ray and the degree of field ordering \citep{tsunetoe_2020_m87pol}. Semi-analytic radiatively inefficient accretion flow (RIAF) models reach a similar conclusion by varying source ingredients one at a time: magnetic geometry and radial inflow drive the largest changes in the polarized image, and reproducing the observed polarization generally requires more structure than is present in the simplest Faraday-thin disk models \citep{saurabh_2025_m87semianalytic}.

A broader M87*-focused review similarly reported that the spin of the black hole remains highly uncertain, even after considering non-polarimetric data \citep{hada_2024_m87review}. That review further agrees that direct photon-ring spin constraints likely require baselines beyond the ground EHT, and that radial evolution of $\arg(\beta_2)$ near the inner shadow is a key future test of spin-powered energy extraction. On larger scales, jet measurements similarly constrain the same central engine although they do not measure the spin: kinematic models of the jet infer the presence of a slowly rotating magnetosphere with field-line angular velocity of order $c/(100\,r_g)$--$c/(70\,r_g)$ \citep{kino_2022_m87slowmag}; multifrequency jet polarimetry finds a helical magnetic field in the acceleration--collimation zone \citep{park_2025_m87helix}; and ALMA Band 7 polarimetry measures linear-polarization fractions of order $1$--$17\%$, RMs exceeding $10^5\,{\rm rad\,m^{-2}}$, and RM gradients or sign reversals suggestive of helical fields \citep{goddi_2025_m87jetalma}. Taken together, these jet-scale measurements constrain the projected jet-axis orientation, the spatial field winding and topology of the magnetic field, and field-line angular velocity $\Omega_F$. Interpreting these quantities still requires a magnetic-field model, however, and the resulting constraints do not provide an independent precision measurement of $\bhspin$.

\subsection{Sgr A*}

The black hole at the center of our own galaxy, Sgr A*, evolves approximately $1500$ times more rapidly than M87*, so the horizon-scale structure changes on minute-to-hour timescales. The EHT total-intensity data measure a ring diameter of $51.8\pm2.3\,\mu{\rm as}$, and model comparisons against numerical simulations disfavor high inclinations, nonspinning black holes, and retrograde disks \citep{eht_sgra_1,eht_sgra_5}, with the constraints coming from a combination of ring size and geometry, the total millimeter flux and size at lower frequencies, near-infrared and X-ray flux limits, as well as temporal variability measures in the total light curve and the structure of the image. Since no fiducial EHT model passes all EHT Paper V constraints, the total-intensity comparison is best read as a kind of filter on source models rather than a spin constraint. The 2017 EHT polarimetric image provides stronger constraints on the source model than the total intensity data alone, inferring a thick polarized ring, a coherent spiral EVPA pattern, a peak linear-polarization fraction near $40\%$, and a less robust circular-polarization dipole at roughly the few-to-ten percent level \citep{eht_sgra_7}. The EHT analysis concludes that the large resolved polarization fraction of $24$--$28\%$ with peaks near $40\%$ is most naturally explained by strongly magnetized models because weakly magnetized or Faraday-thick models tend to depolarize the image too much \citep{eht_sgra_8}.

EHT Paper VIII provides a concrete spin claim only after making a choice for how to model where the Faraday rotation is produced. If most of the mean RM comes from a relatively stable screen outside the emitting region, the observed EVPAs can be derotated, and the handedness of $\beta_2$ selects clockwise models; in the Paper VIII library, only the model with a strongly magnetized flow, $\bhspin=0.94$, $R_{\rm high}=160$, and inclination $i=150^\circ$ satisfies all constraints \citep{eht_sgra_8}. If instead the Faraday rotation happens within the source, the EVPA handedness is counterclockwise and no sampled model passes all constraints. The former treatment, which assumes an external screen, is motivated by and consistent with the observed persistent sign of the RM and with independent inference of clockwise motion \citep{eht_sgra_8,wielgus_2022_sgrapolalma,ricarte_2025_sgrapolcurves}, but it is not guaranteed. Paper VIII estimates that at least $97\%$ of the measured RM must be external, while lower-frequency ALMA data argue for substantial Faraday rotation inside the compact source \citep{eht_sgra_8,wielgus_2024_internalfaraday}. Assuming the same Faraday and emission assumptions, the BHP II model favors low-to-intermediate spins for Sgr A* and disfavors high-spin configurations \citep{wong_2026_bhp2}.

\vspace{0.5em}

Time-domain polarimetry supplies an independent handle on the handedness and orientation of the emitting plasma. ALMA full-Stokes light curves from the 2017 EHT campaign show $Q$--$U$ loops and EVPA rotation on a timescale of about 70 minutes; interpreting this motion as produced by a hotspot infers low-inclination, clockwise equatorial motion at a radius of order $10\,r_g$, a projected angular-momentum axis near $60^\circ$ east of north up to a $180^\circ$ ambiguity, and a weak prograde-spin hint that depends on the Keplerian hot-spot model \citep{wielgus_2022_sgrapolalma}. A later analysis of the 230 GHz polarized light curve finds clockwise motion with a clockwise fraction $0.65\pm0.09$ and favors face-on, clockwise, strongly magnetized models \citep{ricarte_2025_sgrapolcurves}. Because the loop handedness, period, and EVPA evolution test flow orientation and time-dependent transfer, these results check the geometry used in the static polarimetric arguments without relying on the derotated EHT EVPA pattern. They support the clockwise geometry selected by the external-screen interpretation considered in EHT Paper VIII and used as an input for the BHP II analysis.

Near-infrared GRAVITY flare astrometry and polarimetry can also be used as a test of handedness and geometry at the shorter wavelength observed by the GRAVITY instrument. The flares they record show clockwise centroid motion and rotating polarization from radii of a few to about ten $r_g$, supporting a low-to-moderate inclination and a dynamically important poloidal field \citep{gravity_2018_orbitisco,jimenezrosales_2020_gravitybfields,gravity_2023_nirpolarimetry}. More detailed hot-spot, flare-polarimetry, and orbital-tomography models provide additional information about the orientation of the emitting plasma, although their measurement of the spin remains tied to the assumed orbit and emission prescription. Millimeter $Q$--$U$ loop fits can infer inclinations near $155^\circ$--$160^\circ$, periods near 90 minutes, radii of order $9$--$12\,r_g$, and $\bhspin>0.8$, while joint near-infrared (NIR) astrometry/polarimetry and 3D polarimetric tomography recover clockwise, low-inclination orbital structure and emphasize that spin remains weakly constrained by present light curves \citep{yfantis_2024_quloops,yfantis_2024_sgra_hotspots,levis_2024_sgra_tomography}. Simulations of MAD accretion disks with flares also reproduce $Q$--$U$ loops and 30-minute to 1-hour polarimetric periodicities, with only mild spin dependence entering through the orbital dynamics \citep{najafiziyazi_2024_sgraflarepol}. Together, these observations and models support spin inference mainly by constraining the source ingredients needed by image-based arguments: plasma motion, inclination, field geometry, and characteristic timescales. Full-Stokes monitoring during the 2018 EHT campaign gives the same warning from another epoch: total intensity can remain relatively quiet while linear and circular polarization vary strongly during a millimeter/X-ray event, so time-domain polarization is first a source-variability constraint rather than a spin diagnostic \citep{albentosaruiz_2025_sgra2018pol}.

Wielgus et al.~used ALMA observations at $85$--$101$ and $212$--$230$ GHz to infer properties of the Faraday-rotating plasma, finding steep depolarization below $\sim150$ GHz, with the mean RM in the higher-frequency band about twice as large as the RM in the lower-frequency band, as well as rapid RM variability. These findings suggest that much of the Faraday rotation is internal to the compact emitting region, with about half of the rotation occurring inside the inner $\sim10\,r_g$ \citep{wielgus_2024_internalfaraday}. For spin inference, this controls whether EVPA derotation can be treated as a foreground correction or must be modeled as part of the source. Analysis of circular polarization data rejects edge-on $i=90^\circ$ models for Sgr A* and, among EHT-favored MAD models, shows a preference for an orientation of the magnetic dipole moment away from the Earth and clockwise rotation in the source \citep{joshi_2024_circularpol,yin_2025_sgracp}. Surveys of numerical simulations that consider broadband spectra, polarization, RM, variability, and NIR flare information already favored strongly magnetized models over weakly magnetized disk--jet models that overproduce Faraday depolarization \citep{dexter_2020_sgrasurvey}. Electron thermodynamics can also change which models survive: two-temperature MAD calculations improve agreement with 230 GHz variability, so Paper VIII-style filtering is sensitive to thermodynamic assumptions as well as to spin and inclination \citep{salas_2025_sgra_twotemp}.

\vspace{0.5em}

Sgr A* provides a useful example of how the spin question changes as new observables are added. Pre-EHT RIAF and millimeter-VLBI analyses used spectra, unresolved polarization, size, and closure quantities to constrain spin, inclination, and position angle within analytic flow families \citep{broderick_2009_sgraparams,broderick_2011_sgraparams,broderick_2016_modelingeht}. GRMHD polarized-transfer fits to submillimeter data likewise found preferred regions of spin, inclination, and position angle, but with broad allowed ranges and substantial radiative-transfer caveats \citep{shcherbakov_2012_sgrapol}. Resolved polarimetry changes the character of the inference: it constrains not only image size and orientation, but also magnetization, Faraday structure, magnetic polarity, clockwise accretion-flow motion, and likely viewing geometry. The spread between older low-spin source-model priors, the high-spin Paper VIII external-screen solution, and the low-to-intermediate-spin BHP II magnetic-pitch interpretation should therefore be read less as a set of competing spin measurements than as a map of the source assumptions still to be fixed. The path forward is to connect these measured ingredients through a consistent model of both the emitting plasma and the Faraday-active material along the line of sight, so that conditional spin interpretations can sharpen into more robust statements about spin magnitude or spin sense.

\subsection{Other sources}

The jet-base diagnostics described in Section~\ref{sec:connection} suggest a path for spin inference beyond M87* and Sgr A*, and other VLBI sources are beginning to supply the ingredients required for that extension. In these systems, polarimetry typically samples jet material well outside of the horizon-scale emission region, and the data can be used to constrain ordered magnetic fields, transverse polarization and rotation-measure gradients, core shifts, EVPA evolution, and wavelength-dependent Faraday rotation or depolarization by the surrounding accretion flow or jet sheath \citep{park_algaba_2022_agnjetpol}. These observables test the magnetic topology, acceleration and collimation structure, source orientation, propagation environment, and field-line-rotation assumptions that enter jet-based spin diagnostics, including light-cylinder polarimetric forecasts \citep{gelles_2025_spinjetpol,gelles_2026_offaxisjetpolspin}. The central challenge is therefore to distinguish what VLBI polarimetry measures directly about the jet and propagation environment from the additional, model-dependent step that interprets those measurements as constraints on spin magnitude, spin sense, or magnetospheric angular velocity.

Several systems show how close this bridge already is. In NGC 315, VLBI jet acceleration and collimation data and modeling infer a field line angular velocity $c/(800\,r_g)\lesssim\Omega_F\lesssim c/(500\,r_g)$ and, after adopting $0.5\lesssim\bhspin\lesssim1$, a horizon magnetic field of order $5\times10^3$--$2\times10^4$ G; here, the magnitude of the spin is an input, but the inferred $\Omega_F$ is already a concrete magnetosphere constraint that can be used to support future spin inference \citep{kino_2024_ngc315spin}. In 3C~273 and CTA~102, transverse rotation-measure gradients and polarization patterns corrected for Faraday rotation have been used to infer helical or toroidal jet fields; in CTA~102, similar information has also been combined with data on circular polarization to infer an apparent rotation sense given a particular model for the magnetic field in the jet \citep{asada_2002_helical_3c273,gabuzda_2018_bfieldrotation,casadio_2019_cta102,lisakov_2021_3c273}. EHT and high-frequency VLBI observations of J1924$-$2914 and NRAO~530 now also resolve ordered polarization in compact blazar jet bases, enabling tests of the field geometry and Faraday environment \citep{issaoun_2022_j1924-2914,jorstad_2023_nrao530,lisakov_2025_nrao530}. Centaurus A has also been resolved by the EHT at 228 GHz as an edge-brightened jet/counterjet system, with simultaneous ALMA linear polarization constrained below about $0.15\%$ on larger scales, making it a promising future target for resolved polarimetric tests of a spin-linked spine--sheath picture \citep{janssen_2021_cena}. Sub-parsec 22 GHz VLBI polarimetry of 3C~84/NGC~1275 detects limb-brightened polarized emission and a bimodal EVPA structure consistent with a mildly relativistic spine--sheath jet threaded by a predominantly toroidal magnetic field, adding a constraint on magnetic topology for the same family of models \citep{paraschos_2024_3c84}. Together these sources are beginning to assemble the information needed to ask whether diagnostics calibrated in M87* and Sgr A* might carry over to a broader jet population.

\section{Outlook}
\label{sec:outlook}

Horizon-scale polarimetry has made black hole spin inference a more concrete problem. The observed Stokes structure constrains the polarized image, the ordered component of the near-horizon magnetic field, Faraday propagation effects, the magnetic flux state in the accretion flow, source variability, and elements of the relationship between the compact source and the jet. While these are not measurements of $\bhspin$ by themselves, they are spin-sensitive source properties, and a particular spin interpretation therefore depends on how the Kerr geometry, magnetized plasma, emission geometry, and polarized transfer are connected within a common source model.

The current source-by-source picture is correspondingly uneven but informative. For M87*, the strongest polarimetric statements are about ordered magnetic fields, strong magnetization, variability, and the connection to the larger jet. The spin-axis orientation can be tied to the observed jet if the black hole, disk, and jet are aligned or anti-aligned, and several simulation-library and semi-analytic analyses find spin-sensitive trends. The spin magnitude, however, remains tied to assumptions about jet power, electron thermodynamics, Faraday effects, emission geometry, and variability. For Sgr A*, polarimetry constrains magnetization, clockwise source motion, magnetic polarity, Faraday structure, and time-dependent emission. Its spin interpretation remains sensitive to whether Faraday rotation is treated as an external screen or as part of the compact source, and to how the observed EVPA pitch is connected to the magnetic field in the emitting plasma. 

Other VLBI sources extend the same logic to compact jets, where polarimetry constrains field topology, rotation-measure structure, field-line rotation, and viewing geometry before those quantities can be translated into spin magnitude or spin sense. Looking ahead, spin-sensitive polarimetric tests should increasingly reach beyond M87* and Sgr A*. With improved EHT/ngEHT sensitivity, broader frequency coverage, and larger source samples, jet polarimetry and near-horizon image- and visibility-domain $\beta_2$-type measurements could extend these tests to many more systems \citep{pesce_2022_expectationsngeht}. In that regime, consilience would mean consistency across a population as well as within individual sources. Even if any one object is less constrained than M87* or Sgr A*, measurements of compact polarized structure, jet orientation, Faraday behavior, and variability can still be compared across the sample to search for systematic patterns in spin-sensitive magnetic structure \citep{palumbo_2020_beta2,palumbo_2025_spinconstraints}.  Table~\ref{tab:spinpol_inventory} summarizes the main inference targets, their present caveats, and the corresponding paths forward.

\begin{table}[H]
\caption{Summary of spin-related quantities, the polarimetric information that can constrain them, and the next steps for the inference problem. \label{tab:spinpol_inventory}}
\footnotesize
\begin{tabularx}{\textwidth}{LLLL}
\toprule
\textbf{target} & \textbf{polarimetric information} & \textbf{status and caveats} & \textbf{path forward} \\
\midrule
spin magnitude & simulation library trends; horizon and photon-ring signatures; jet light-cylinder scale & model dependence and astrophysical uncertainties, including magnetic flux and field geometry, inclination, electron thermodynamics, emissivity structure, and the importance of Faraday effects & broader image libraries and modeling; long-baseline polarimetry; resolved jet-base structure \vspace{0.8em} \\
spin-axis orientation & image inclination and asymmetry; resolved jet orientation; multi-scale source geometry & most plausible when tied to an independently observed jet, as in M87*, but this assumes disk/jet/spin alignment & joint horizon-scale and jet modeling across the disk--jet interface; multi-epoch tests of persistent orientation \vspace{1.8em} \\
spin sign and magnetic handedness & EVPA helicity; circular polarization; Faraday-rotation structure; helical jet fields & assumption-heavy, requiring field polarity, the sense of fluid rotation or the approaching side of the jet, where Faraday rotation/conversion occurs, and the relation between disk, jet, and spin axes & multifrequency linear and circular polarimetry; RM mapping; calibrated jet/horizon comparisons \vspace{1.8em} \\
magnetic flux state & linear-polarization morphology; $\beta_m$; jet power; numerical model comparison & relatively well constrained compared to $\bhspin$, but is a source property rather than a spin measurement & EHT/ngEHT model comparison; synthetic-data tests; simultaneous jet-power constraints \vspace{0.8em} \\
field polarity and plasma content & circular polarization; frequency-dependent EVPA structure & promising for magnetic polarity and low-temperature electrons, but sensitive to calibration and transfer properties & better Stokes $V$ calibration; multifrequency transfer modeling; ALMA/EHT/ngEHT comparisons \vspace{1.8em} \\
time-dependent geometry & polarized light curves; $Q$--$U$ loops; hotspot motion & promising for inclination, rotation sense, and characteristic timescales, but current models often idealize the velocity field and surrounding flow & dynamic imaging; simultaneous multiwavelength monitoring; hierarchical inference across epochs \vspace{0.8em} \\
horizon and photon-ring behavior & (near-)horizon EVPA trends; long-baseline polarimetric phases; subring structure & cleaner geometrically, but faint, demagnified, and contaminated by direct or foreground emission & space VLBI/BHEX-like baselines; higher-frequency arrays; closure quantities and visibility-domain polarization ratios with reduced gain sensitivity \vspace{0.8em} \\
\bottomrule
\end{tabularx}
\end{table}

For individual sources, clearer spin inference will require progress along the full chain from polarimetric data to physical interpretation: interferometric measurement, reconstruction or modeling of polarized image structure, time-dependent plasma dynamics and polarized transfer, and finally the connection with $\bhspin$. The items below summarize major sources of uncertainty.

\begin{enumerate}[leftmargin=*]
    \item \emph{Time variability.} Temporal evolution of the emitting plasma can muddy the signatures of persistent geometry with transient stochastic features.  This is important for the rapidly varying Sgr A* as well as for M87*, where multi-epoch data show direct evidence that the polarized morphology can change with time in temporally resolved observations. Dynamic imaging, multi-epoch modeling, and simultaneous multiwavelength monitoring can help distinguish persistent source geometry from turbulent variability. More organized time-domain signatures, such as $Q$--$U$ loops, total-intensity hotspot motion, coherent EVPA evolution, or spin-sensitive variability timescales, may then serve as model-discriminating observables when the source model connects flaring and quiescent structure.
    \item \emph{Propagation and transfer.} Faraday rotation, Faraday conversion, opacity, and scattering can change EVPAs, alter $\beta_m$, depolarize ordered emission, and generate Stokes $V$. Multifrequency full-Stokes polarimetry, especially when observations at $86$, $230$, and $345\,{\rm GHz}$ can be compared, is the main route to separating intrinsic source structure from propagation effects.
    \item \emph{Source-model priors.} Numerical simulation libraries are produced with choices about the disk magnetic flux and polarity, tilt, fluid thermodynamics, particle content, and observer inclination. If simulations remain a major point of comparison for the data, broader libraries and reduced models should vary these ingredients and explore the importance of the initial condition, simulation duration, resolution, and other numerical modeling choices.
    \item \emph{Jet connection.} Thus far, translating observations into constraints on jet structure, disk/jet/black hole alignment, spin, or geometric tracers like the pitch of the magnetic field or the location of the light cylinder relies on assumptions about the relationship among the source components, the propagation screen, and the dynamics and structure of the jet. The disk--jet interface is a particularly important part of this problem because it controls how horizon-threading flux, disk-launched material, mass loading, and sheath emission connect the compact polarized image to the resolved outflow. These assumptions can be tested by connecting horizon-scale and jet-scale polarimetry within a common framework for the magnetic field and by tracing the magnetic field from the large-scale jet down to horizon scales.
    \item \emph{Access to near-horizon and photon-ring signatures.} Spin signatures arising from emission very near the horizon or from strong lensing along the photon ring are cleaner in principle, but accessing them is observationally and astrophysically difficult: they are faint, redshifted, demagnified, blended with direct emission, and sensitive to foreground or off-equatorial plasma. Higher-frequency arrays and space-VLBI concepts such as BHEX will be needed to reach regimes where lensing, horizon boundary conditions, and near-horizon field structure can be tested more directly. These tests include separating direct and lensed image orders, measuring radial trends toward a horizon-set EVPA, and detecting visibility-domain photon-ring signatures.
    \item \emph{Data and reconstruction.} Even when a proposed feature is accessible in theory, the actual observing array must be able to measure the relevant signal. Additional telescopes, better baseline coverage, higher sensitivity, broader frequency coverage, and improved polarization calibration determine which visibility-domain constraints are realistically accessible. Image reconstruction then determines how those constraints appear in images and summary statistics. Direct visibility-domain analysis and synthetic-observation tests will keep proposed spin signatures tied to the data.
\end{enumerate}

Progress should be judged by consistency across observables with complementary systematics. A spin interpretation becomes more compelling when EVPA morphology, $\beta_m$, Stokes $V$, Faraday behavior, time variability, jet structure, and any accessible horizon or photon-ring signature converge on the same source geometry and magnetic field configuration. This kind of consilience is stronger than agreement obtained by changing the Faraday prescription, field polarity, thermodynamics, or source geometry for each observable separately. Non-detections and failures of proposed spin signatures are also useful when they identify where variability, transfer, plasma thermodynamics, or emission geometry erases an otherwise spin-sensitive imprint. Future observing and modeling programs should prioritize combinations of polarimetric diagnostics with complementary systematics and test whether they support a common spin-dependent interpretation across independent data products and plausible source models.

\vspace{6pt}

%%%%%%%%%%%%%%%%%%%%%%%%%%%%%%%%%%%%%%%%%%
% \authorcontributions{For research articles with several authors, a short paragraph specifying their individual contributions must be provided. The following statements should be used ``Conceptualization, X.X. and Y.Y.; methodology, X.X.; software, X.X.; validation, X.X., Y.Y. and Z.Z.; formal analysis, X.X.; investigation, X.X.; resources, X.X.; data curation, X.X.; writing---original draft preparation, X.X.; writing---review and editing, X.X.; visualization, X.X.; supervision, X.X.; project administration, X.X.; funding acquisition, Y.Y. All authors have read and agreed to the published version of the manuscript.'', please turn to the  \href{http://img.mdpi.org/data/contributor-role-instruction.pdf}{CRediT taxonomy} for the term explanation. Authorship must be limited to those who have contributed substantially to the work~reported.}

\funding{G.N.W.~gratefully acknowledges support from the Institute for Advanced Study, through the Taplin Fellowship, and from the Princeton Gravity Initiative. A.C.~was supported by the Villum Fonden grant No. 82533. Z.G.~was supported by a NSF graduate research fellowship. D.C.M.P.~acknowledges financial support from the National Science Foundation (AST-2307887). This work was supported by the Black Hole Initiative, which is funded by grants from the John Templeton Foundation (Grant 62286) and the Gordon and Betty Moore Foundation (Grant GBMF-8273) - although the opinions expressed in this work are those of the author(s) and do not necessarily reflect the views of these Foundations.}

\acknowledgments{We are grateful to the organizers of the ``Taking Spin Measurements for a Spin: Recent Progress on Black Hole Spin Measurements Across the Electromagnetic and Gravitational Spectra'' Workshop at Wake Forest University in September 2025 for triggering insightful discussions that led to the preparation of this document. During the preparation of this manuscript, ChatGPT v5.5 was used to design parts of Figure 2 and for proofreading. The authors have reviewed and edited the output and take full responsibility for the content of this publication.}

\conflictsofinterest{The authors declare no conflicts of interest.}

\abbreviations{Abbreviations}{
The following abbreviations are used in this manuscript:
\\

\noindent
\begin{tabular}{@{}ll}
ALMA & Atacama Large Millimeter/submillimeter Array\\
BHEX & Black Hole Explorer\\
BHP & Black Hole Polarimetry\\
BZ & Blandford--Znajek\\
EHT & Event Horizon Telescope\\
EVPA & Electric vector position angle\\
GRMHD & General relativistic magnetohydrodynamics\\
IAU & International Astronomical Union\\
ISCO & Innermost stable circular orbit\\
MAD & Magnetically arrested disk\\
ngEHT & Next-generation Event Horizon Telescope\\
NIR & Near-infrared\\
PWP & Palumbo--Wong--Prather\\
RIAF & Radiatively inefficient accretion flow\\
RM & Rotation measure\\
SANE & Standard and normal evolution\\
Sgr A* & Sagittarius A*\\
VLBI & Very long baseline interferometry
\end{tabular}
}

%%%%%%%%%%%%%%%%%%%%%%%%%%%%%%%%%%%%%%%%%%
%\isPreprints{}{% This command is only used for ``preprints''.
\begin{adjustwidth}{-\extralength}{0cm}
%} % If the paper is ``preprints'', please uncomment this parenthesis.
%\printendnotes[custom] % Un-comment to print a list of endnotes

\reftitle{References}

\bibliography{bibstrings,references}

\PublishersNote{}
%\isPreprints{}{% This command is only used for ``preprints''.
\end{adjustwidth}
%} % If the paper is ``preprints'', please uncomment this parenthesis.
\end{document}